\documentclass[12pt,english,floatfix,superscriptaddress,aps,prd,preprint,showkeys,nofootinbib]{revtex4}
\usepackage{amsmath}
\usepackage{amssymb}
\usepackage{amsbsy}
\usepackage{amsfonts}
\usepackage{amsopn}
\usepackage{amstext}
\usepackage{graphicx}
\usepackage[english]{babel}
\usepackage{color}
\usepackage{slashed}
\usepackage{esint}
\usepackage[dvips]{epsfig}
\usepackage[dvips]{graphicx}
\usepackage{float}
\usepackage{units}
\usepackage{textcomp}
\usepackage{wasysym}
\usepackage{hyperref}
\usepackage{slashed}

\newcommand{\be}{\begin{align}}
	\newcommand{\ee}{\end{align}}
\newcommand{\bea}{\begin{eqnarray}}
	\newcommand{\eea}{\end{eqnarray}}
\newcommand{\hf}{\frac{1}{2}}

\newcommand{\sech}{\text{sech}}
\newcommand{\e}{\textrm{e}}

\begin{document}

\title{Geometrical deformation of brane matter field within $f(R,Q,P)$ gravity}

\author{Fernando M. Belchior}
\email{belchior@fisica.ufc.br}
\affiliation{Departamento de F\'isica, Universidade Federal do Cear\'a,\\ Campus do Pici, 60455-760, Fortaleza, Cear\'a, Brazil.}
\affiliation{Departamento de F\'isica, Universidade Federal da Para\'iba,
 58051-970, João Pessoa, Para\'iba, Brazil}

\author{Roberto V. Maluf}
\email{r.v.maluf@fisica.ufc.br}
\affiliation{Departamento de F\'isica, Universidade Federal do Cear\'a,\\ Campus do Pici, 60455-760, Fortaleza, Cear\'a, Brazil.}

\author{Albert Yu. Petrov}
\email{petrov@fisica.ufpb.br}
\affiliation{Departamento de F\'isica, Universidade Federal da Para\'iba,
 58051-970, João Pessoa, Para\'iba, Brazil}

\author{Paulo J. Porf\'irio}
\email{pporfirio@fisica.ufpb.br}
\affiliation{Departamento de F\'isica, Universidade Federal da Para\'iba,
 58051-970, João Pessoa, Para\'iba, Brazil}

\begin{abstract}
In the context of braneworld scenarios, the real scalar field plays a crucial role by providing thickness for brane. In this work, we investigate a codimension-one thick brane within the framework of $f(R,Q,P)$ gravity, where $R$ represents the curvature scalar, while $Q=R_{\mu\nu}R^{\mu\nu}$ and $P=R_{\mu\nu\alpha\beta}R^{\mu\nu\alpha\beta}$ are quadratic geometric invariants. Our interest is to investigate the influence of these invariants on the scalar field solution and the energy density of the brane. Furthermore, we analyze the localization of spin $1/2$ fermions and the gravitino by employing a Yukawa-like coupling with the scalar field background. Such a coupling is able to produce a normalizable zero-mode for both fields. 

\end{abstract}

\maketitle

%%%%%%%%%%%%%%%%%%%%%%%%%%%%%%%%%%%%%%%%%%%%%%%%%%%
\section{Introduction}

Nearly a century ago, Kaluza and Klein published a remarkable work in which the hypothesis of extra dimension was proposed for the first time, being motivated by an idea to combine gravity and electrodynamics in a single unified theory \cite{Kaluza:1921tu, Klein:1926tv}. Further, various estimations for the size of compact extra dimensions were done, for a review see e.g.\cite{Antoniadis:1998ig}. Proposed in 1999, the so-called Randall-Sundrum (RS) model introduced another approach to extra dimensions \cite{rs,rs2}, aiming to solve the hierarchy problem, a persistent question in theoretical physics for many years, which, in particular, discusses the weakness of gravitational interaction compared to other interactions. In the RS model, the metric is described by a function of the extra spatial coordinate introduced in such a manner that gravity could escape to the extra dimension and is considerably weak in our universe (3-brane) \cite{Garriga:1999yh}. On the other hand, the fields responsible for the other fundamental interactions would be trapped on a 3-brane.

 Generalization of the RS theory, also known as braneworld models, have been considerably investigated over the past two decades. It is worth to emphasize especially the role played by thick branes which represent a direct extension of the RS model, where one introduces a real scalar field allowing for a more general form for the warp factor of the RS metric, providing thickness for the brane \cite{Gremm1999, Bazeia2004, Bazeia:2020qxr, Dzhunushaliev2009, Charmousis2001, Arias2002, Barcelo2003, CastilloFelisola2004, Dutra2014, Koley:2004at}. Another important feature of thick branes is the appearance of their internal structure \cite{Bazeia2004, Bertolami:2007dt}. In (1+1) dimensions, kink-like structures emerge from scalar field dynamics \cite{Dorey:2011yw, Campbell:1986nu, Gani:1998jb, Mussardo:2004rw, Gani:2017yla, Lima:2022chw, Andrade:2023jvf}. Such structures possess $Z_2$ symmetry and interpolate between the vacuum expectation values of the theory, which stem from a spontaneous symmetry-breaking mechanism. It is worth noting that spontaneous symmetry breaking also leads to nontrivial topological structures such as vortices \cite{Lima:2021amb, Lima:2021gsl, Lima:2023str, Casana:2017yrs} and monopoles \cite{Benes:2023nsr}.

Within the thick brane scenarios, in addition to the scalar field dynamics, one can modify the geometrical sector by considering alternative gravity models like, e.g., $f(R)$ and $f(R,T)$ gravities \cite{Afonso:2007gc, Nascimento:2018sir, Rosa1, Rosa2, HoffdaSilva:2011si, Carames:2012gr, Gu:2014ssa, Bazeia:2013uva, Bazeia:2015oqa, Zhong:2015pta, Yu:2015wma, Guo:2019vvm, Carlos2024, Correa:2015qma,daSilva:2017jbx,Rosa3,Rosa4}, where $R$ represents the curvature scalar and $T$ the trace of momentum-energy tensor. Other theories investigated within the brane context were cubic gravity \cite{Lessa:2023thi}, mimetic gravity \cite{Lobao:2022inw}, bimetric gravity \cite{Bertolami:2003ui}, a scalar-tensor model known as Horndeski gravity \cite{Brito:2018pwe}, bumblebee gravity \cite{Marques:2023suh} and scalar fields with generalized dynamics \cite{Menezes}.

A pivotal issue of thick branes is analyzing the localization of Standard Model fields on the brane. Such an investigation is carried out by considering the five-dimensional action for a specific field. Then one employs a Kaluza-Klein decomposition that allows us to obtain the zero-mode (massless mode) of the field and verify whether it is indeed localized on the brane. In the case of gauge fields \cite{Bertolami:2007dt, Cruz:2010zz, Freitas:2018iil} and fermions \cite{Liu:2009ve, Melfo:2006hh, Liu:2008pi, Liu:2009dw, Chumbes:2010xg, Almeida:2009jc, Cruz:2011ru}, an additional coupling is necessary to yield a normalizable zero mode. Over the past years, several articles have been addressing this subject, most of them in the context of general relativity \cite{Bertolami:2007dt, Cruz:2010zz, Freitas:2018iil, Liu:2009ve, Melfo:2006hh, Liu:2008pi, Liu:2009dw, Chumbes:2010xg, Almeida:2009jc, Cruz:2011ru}.

In this work, we are interested in studying a five-dimensional thick brane in the $f(R, Q, P)$ gravity coupled to a single scalar field with standard dynamics. The $f(R, Q, P)$  model, whose action depends on scalar curvature $R$ and squares of Riemann and Ricci tensors, $P$ and $Q$ respectively, has been originally introduced in \cite{Bogdanos:2009tn}.  Its important particular case is the $f(R, G)$ gravity, whose action involves the Gauss-Bonnet term $G$. This model was considered mostly within the cosmological context, see f.e. \cite{DeFelice:2010sh, DeLaurentis:2015fea, Benetti:2018zhv, Odintsov:2018nch}. One chooses a suitable warp factor to find matter field solutions for our brane system. Furthermore, the localization of spin $1/2$ (Dirac fermions) and spin $3/2$ (gravitino) zero-modes are investigated. The study of these fields is a relevant question to address in the braneworld models. As one knows, the spin 1/2 spinor describes the ordinary matter, while the spin 3/2 gravitino field (superpartner of the graviton) arises in the supergravity context, being regarded as a possible candidate for dark matter \cite{Buchmuller:2007ui, Steffen:2006hw}.

The outline of this work is the following: In section (\ref{s2}), it is obtained the equations of motions for $f(R, Q, P)$ gravity. In section (\ref{s3}), we construct the braneworld scenario by considering a single scalar field as the matter source. In section (\ref{s4}), it is verified whether the brane system is capable of trapping the fermion zero-mode, and in section (\ref{s5}), the similar study is performed for the Rarita-Schwinger field. Our final discussion, as well as future perspectives, are presented in section (\ref{s6}).
%%%%%%%%%%%%%%%%%%%%%%%%%%%%%%%%%%%%%%%%%%%%%%%%%%%%%%%%%%%%%%%%%%%%%%%%%%%%%%
\section{$f(R,Q,P)$ gravity}\label{s2}

As it is well known, the general relativity (GR) is capable of explaining the geometrical effects of gravity and treating a wide range of phenomena at great scales. However, some open issues such as problems of dark energy and formulating a consistent quantum description demand possible modifications of GR. In this context, modified gravity models have gained increasing room. A simple way to construct such models is to assume that the gravitational action depends not only on the Ricci scalar $R$ but on a generic function of $R$, $f(R)$. Other more robust modifications involve quadratic invariants, that is, the square of Ricci tensor $Q=R_{\mu\nu}R^{\mu\nu}$, the square of Riemann tensor $P=R_{\mu\nu\alpha\beta}R^{\mu\nu\alpha\beta}$ or the Gauss-Bonnet term $G=R^2-4R_{\mu\nu}R^{\mu\nu}+R_{\mu\nu\alpha\beta}R^{\mu\nu\alpha\beta}$. As a starting point for our investigation,  let us define the five-dimensional action for $f(R,Q,P)$ gravity (see e.g. \cite{Nascimento:2023zok})
\begin{align}\label{gaction}
     S=\int d^5x\sqrt{-g}\bigg(\frac{1}{4}f(R,Q,P)+\mathcal{L}_m\bigg),
\end{align}
where $\sqrt{-g}$ is the determinant of the metric $g_{MN}$, $f(R,Q,P)$ represents a generic function of Ricci scalar $R$ and higher-order derivative curvature invariants $Q$ and $P$. Moreover, $\mathcal{L}_m$ is the matter Lagrangian, which will be defined in the next section. Herein, the bulk coordinate indices are denoted by capital Latin index $M=0,\ldots,D-1$. The variation of action (\ref{gaction}) with respect to metric $g_{MN}$ yields the following equation of motion 

\begin{align}
    R_{MN}f_R-\frac{1}{2}g_{MN}f+H_{MN}+K_{MN}=2T_{MN},
\end{align}
where
\begin{align}
  H_{MN}=(g_{MN}\square-\nabla_M \nabla_N)f_R,
\end{align}
\begin{align}
K_{MN}=\square(f_Q R_{MN})+g_{MN} \nabla_A\nabla_B(f_Q R^{AB}) +2f_Q R^A_{(M}R_{N)A}\nonumber\\-2\nabla_A \nabla_{(M}(f_Q R^A_{N)})+f_P R_{ABCM}R^{ABC}\ _N-4\nabla_A \nabla_B(f_P R^A\ _{(MN)}\ ^B).
\end{align}
Above $\square=\nabla^M\nabla_M$, $f_R=\frac{\partial f}{\partial R}$, $f_Q=\frac{\partial f}{\partial Q}$ and $f_P=\frac{\partial f}{\partial P}$. Besides, $T_{MN}$ stands for the matter energy-momentum tensor given explicitly by
\begin{align}
 T_{MN}=-\frac{2}{\sqrt{-g}} \frac{\delta(\sqrt{-g}\mathcal{L}_m)}{\delta g^{MN}}. 
\end{align}
Before proceeding with our study, we must note the fact that higher-order derivative terms potentially lead to ghost-like instabilities that break the unitarity. Similarly to previous works, f.e. \cite{Gama:2017eip, Nascimento:2023zok}, we will circumvent this issue by treating our theoretical model in the context of effective field theories  (for the discussion of this methodology, see \cite{Georgi}). In other words, the effects of higher-curvature terms are Planck-suppressed, so that ghost-like degrees of freedom are neglected at the low energy limit. In the following, we will define the background and the matter sources necessary to perform our study.
%%%%%%%%%%%%%%%%%%%%%%%%%%%%%%%%%%%%%%%%%%%%%%%%%%%%%%%%%%%%%%%%%%%%%%%%%%%%%%%
\section{Deformed thick brane}\label{s3}

In this section, we will present a brief review of the main features of braneworlds space-time in the context of $f(R,Q,P)$ gravity. Adopting the signature of the metric $(-++++)$, the line element of these metrics is given by \cite{rs,rs2}
\begin{align}\label{rsmetric}
    ds^2=\e^{2A(y)}\eta_{\mu\nu}dx^\mu dx^\nu +dy^2,
\end{align}
where $\eta_{\mu\nu}$ (here and further Greek indices label brane coordinates, $\mu=0,..., D-2$) is the Minkowski metric, $\e^{2A(y)}$ is the warp factor which depends only on the extra dimension coordinate $y$. For the braneworld metric (\ref{rsmetric}), the scalars $R$, $Q$ and $P$  can be straightforwardly found to look like:
\begin{eqnarray}
    &R&=-4 (2 A''+5 A'^2),\nonumber\\
    &Q&=20 A''^2+80 A'^4+64 A'^2 A'',\nonumber\\
    &P&=24 A'^4+16 (A''+A'^2)^2.
\end{eqnarray}

\begin{figure}[ht!]
\begin{center}
\begin{tabular}{ccc}
\includegraphics[scale=0.45]{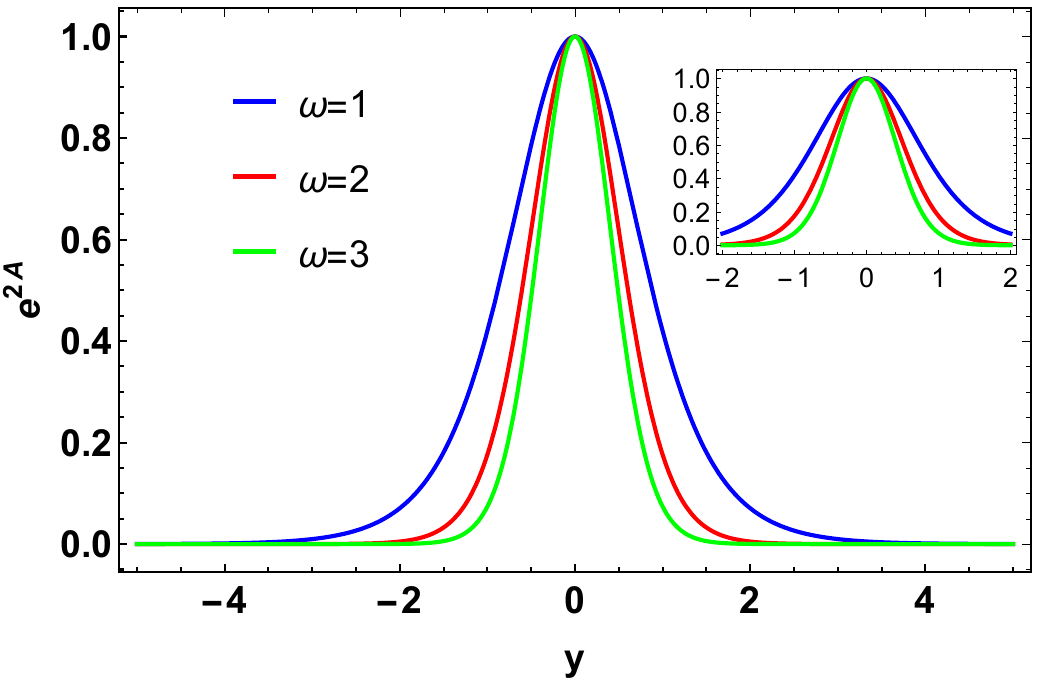} 
\includegraphics[scale=0.45]{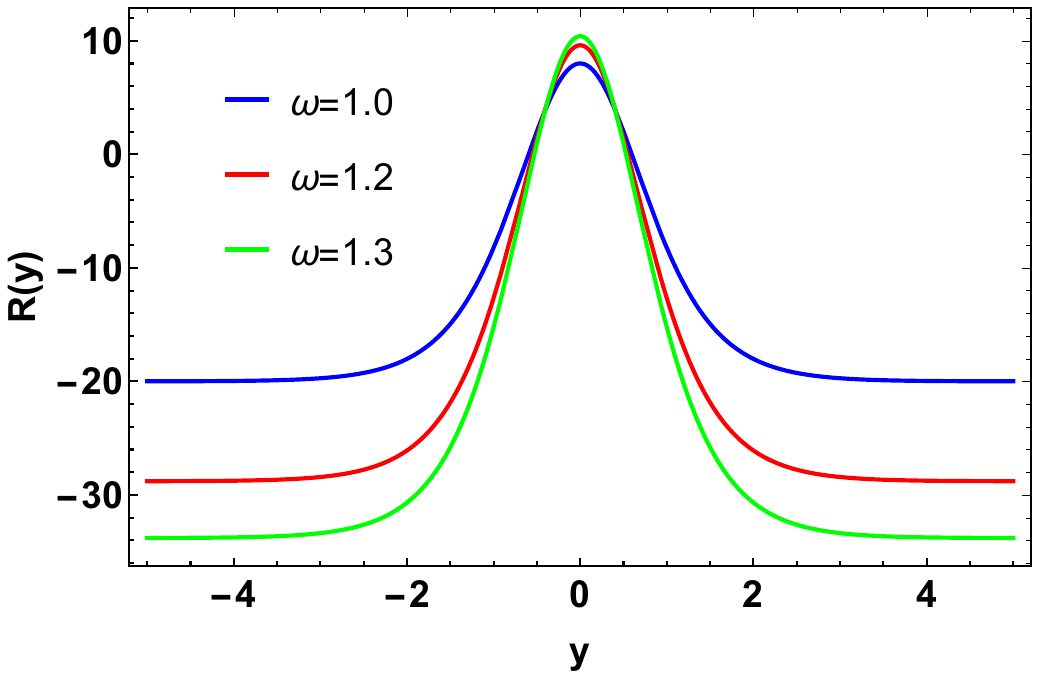}\\
(a)\hspace{7,7cm}(b)\\
\includegraphics[scale=0.45]{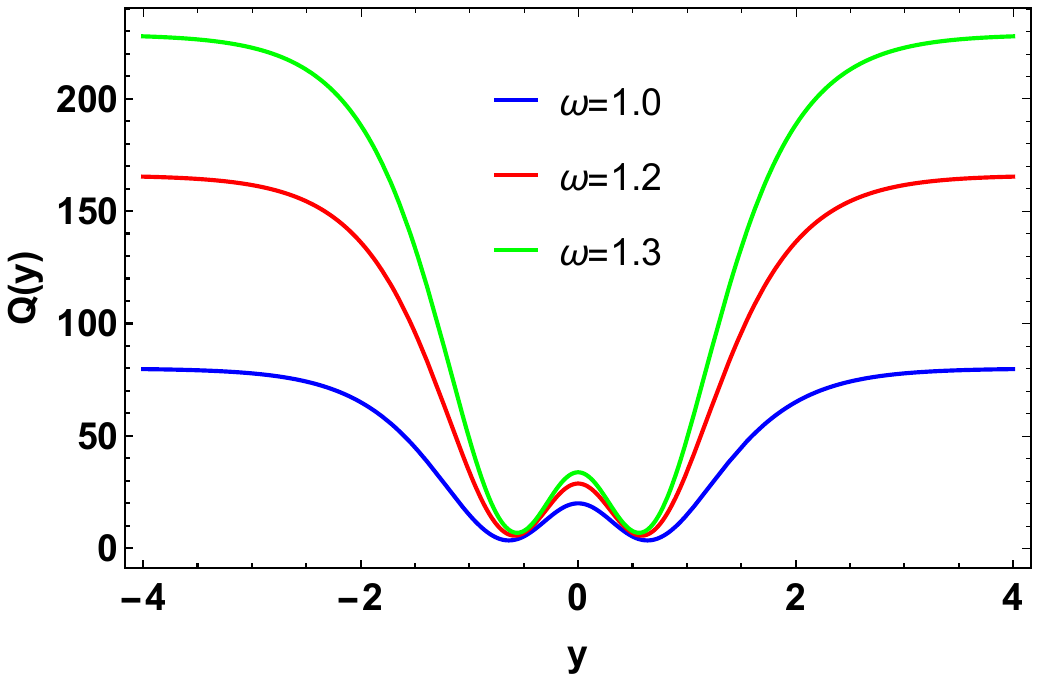} 
\includegraphics[scale=0.45]{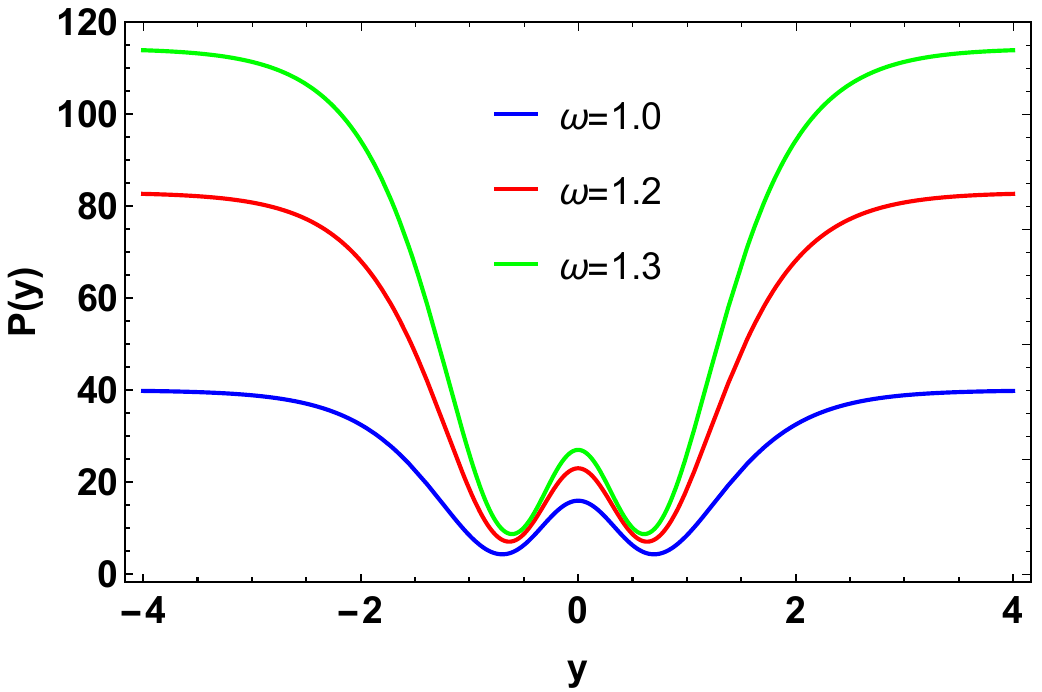}\\
(c)\hspace{7.7cm}(d)
\end{tabular}
\end{center}
\vspace{-0.5cm}
\caption{(a) Warp factor. (b) Ricci scalar. (c) Plot of $Q$. (d) Plot of $P$ 
\label{fig1}}
\end{figure}

Now, we should specify the matter content that will source the braneworld. But firstly,  let us present some physical reasons for our choice. In high-energy physics, spontaneous symmetry breaking plays a pivotal role by giving rise to cosmological-scale structures such as domain walls and cosmic strings. The characteristics of these structures can be investigated within the framework of topological defects. In particular, we can mention especially kink, vortex, and monopole solutions, allowing for the development of analogies between high energy and condensed matter physical theories, the so-called analogue models for gravity \cite{Visser:2001fe}. For instance, the kink-like structures are generated by scalar fields in two-dimensional flat spacetime and described by the Lagrangian below
\begin{align}\label{scalar2}
\mathcal{L}_m=-\frac{1}{2}\partial_c\phi\partial^c\phi-V(\phi),
\end{align}
where $c=0,1$ and $V(\phi)$ represents the potential responsible for spontaneous symmetry breaking. The kink-like configurations actually attract great attention caused by interest to their further generalizations, e.g., double-kink and compacton-like solutions. The double-kink configurations appear, in principle, in theories with three vacuum or using deformation approaches of the standard model, see e.g. Ref. \cite{Dorey:2011yw}. Among some  models allowing for arising of double-kink-like structures one can cite the double-sine-Gordon (DSG), polynomial, and non-polynomial models \cite{Gani:2017yla, Campbell:1986nu, Gani:1998jb, Mussardo:2004rw}. In general, one can also find compacton-like configurations in high-energy physics \cite{Rosenau:1993zz}. However, these configurations appeared for the first time in the context of hydrodynamics. Thus, compacton-like structures have gained prominence due to their strictly localized energy profile \cite{Lima:2021amb}. Recently, the studies on compactons extend from flat spacetime to curved one with applications in cosmology and braneworld scenarios. Moreover, it is still possible to obtain multi-kinks configurations \cite{Andrade:2023jvf}. Following these motivations, let then us assume that a single scalar field is the matter source that generates the thick brane. The matter Lagrangian is given by
\begin{equation}\label{scalar5}
\mathcal{L}_m=-\frac{1}{2}\partial_M\phi\partial^M\phi -V(\phi).   
\end{equation}
The energy-momentum tensor associated to this Lagrangian reads
\begin{equation}
T_{MN}=\partial_M\phi\partial_N\phi+g_{MN}\mathcal{L}_m.    
\end{equation}
The variation of matter action leads to 
\begin{align}
    \frac{1}{\sqrt{-g}}\partial_M(\sqrt{-g}\partial^M\phi)-V_{\phi}=0,
\end{align}
where $V_{\phi}=\frac{dV}{d\phi}$. With the metric in hands, we obtain the following set of equations
\begin{align}
 &5 A^{(4)} f_Q+8 A^{(3)} f_P'+10 A^{(3)} f_Q'+4 A'' f_P''+5 A'' f_Q''+\nonumber\\&20 f_Q A''^2+4 A'^2 F_P''+12 A'^3 f_P'+8 A'^2 f_Q''+24 A'^3 f_Q'-\nonumber\\&32 f_Q A'^4-3 A' f_R'+40 A^{(3)} f_Q A'+40 A' A'' f_P'+56 A' A'' f_Q'+\nonumber\\&56 f_Q A'^2 A''+f_R (A''+4 A'^2)+4 f_P (A^{(4)}+4 A''^2-\nonumber\\&4 A'^4+8 A^{(3)} A'+13 A'^2 A'')-f_R''+\frac{f}{2}=\phi '^2+2V,
\end{align}

\begin{align}
&4 A''^2 (4 f_P+5 f_Q)+4 A'[f_R'- 4 A'^2 (-A' (f_P+2 f_Q)+f_P'+2 f_Q')+\nonumber\\&A'f_R]-4 A^{(3)} A' (4 f_P+5 f_Q)-4 A'' (12 A'^2 (f_P+F_Q)+4A'f_P')+\nonumber\\&20A' f_Q'+f_R)-\frac{f}{2}=\phi '^2-2V,   
\end{align}
and
\begin{align}
\phi^{\prime\prime}+4A^{\prime}\phi^{\prime}=V_{\phi},
\end{align}
where the prime stands for the derivative with respect to the extra dimension $y$. Now, the challenge is solving the set of equations to obtain a scalar field solution. To simplify the problem, we can introduce the standard ansatz for the warp factor (see f.e. \cite{Bazeia:2015oqa}):
\begin{align}
 A(y)=\omega \ln{[\sech{(y)}]}   
\end{align}
It is important to point out that this warp factor fulfills some requirements:

\begin{itemize}
 \item The warp factor must reproduce the RS warp factor far from the brane core which requires  $\lim\limits_{y\rightarrow \infty}e^{2A(y)}\rightarrow 0$.
 \item The warp factor should have a smooth profile near the brane.
  \item It is necessary to ensure that the matter field has $Z_2$ symmetry, then $e^{2A(y)}=e^{2A(-y)}$.
   \item $\int_{-\infty}^{\infty}dy e^{8A(y)}$ must be finite and non-null to ensure a normalizable zero-mode for graviton.
\end{itemize}
In Fig. (\ref{fig1}), we depict the shape of the warp factor as well as the geometrical invariants $R$, $Q$ and $P$. As we can observe, the warp factor indeed displays smooth behavior near the brane, while it tends to zero far from the brane. On the other hand, the curvature scalar $R$ assumes a positive value $8\omega$ at $y=0$. Meanwhile, far from the brane, $R$ approaches a negative constant given by $-20 \omega^2$. Such a result characterizes the $AdS_5$ limit for the bulk. 
Once we have discussed the basics aspects of warped geometry,
our interest now is to understand how the matter field will behave when we modify geometrical parameters. Now, let us study some examples of the function $f(R,Q,P)$. We start with discussing the case of quadratic gravity.

\subsection{Quadratic gravity}
The first model we will analyze is a quadratic gravity, with
\begin{align}
 f(R,Q,P)=R+k_1\, R^2+k_2\, Q + k_3\, P,   
\end{align}
where the parameters $k_{1,2,3}$ control the influence of quadratic invariants.
We obtain the scalar field solution by solving numerically the equation
\begin{align}
\phi^{\prime 2}(y)&=\frac{1}{2} \omega\,  \sech^2(y) \bigg[-4 \tanh ^2(y) (2 k_1 (\omega  (5 \omega +16)+8)+k_2 (2 \omega  (\omega +5)+5)\nonumber\\&+k_3 (\omega  (\omega +8)+4))+2 (4 \omega +1)\, \sech^2(y) (16 k_1+5 k_2+4 k_3)+3\bigg]    
\end{align}

\begin{figure}[ht!]
\begin{center}
\begin{tabular}{ccc}
\includegraphics[scale=0.45]{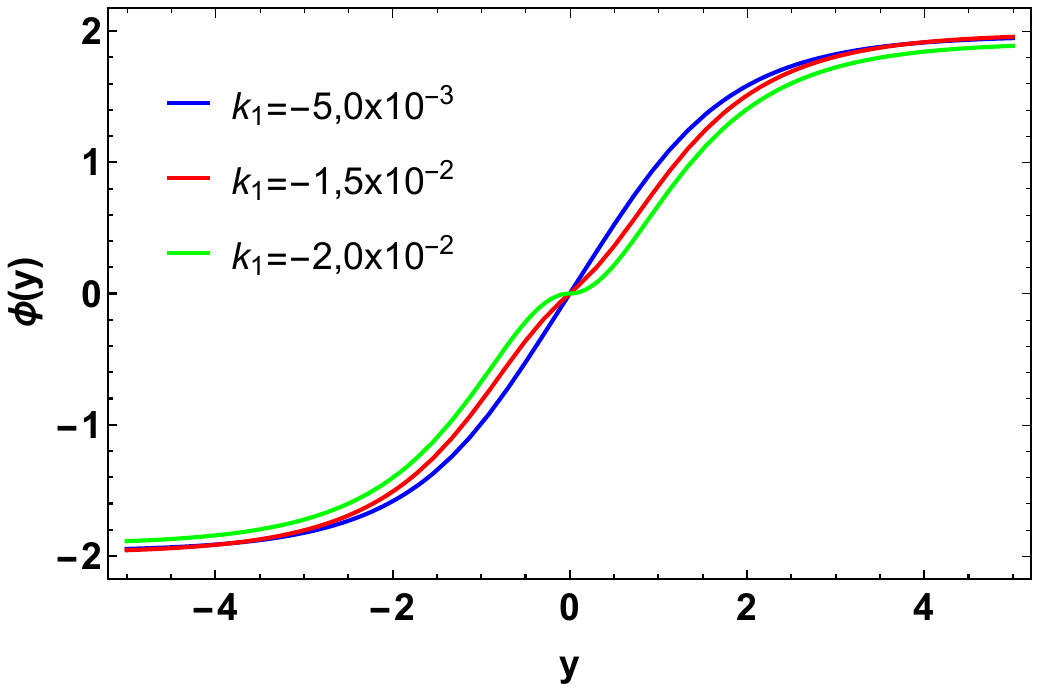} 
\includegraphics[scale=0.45]{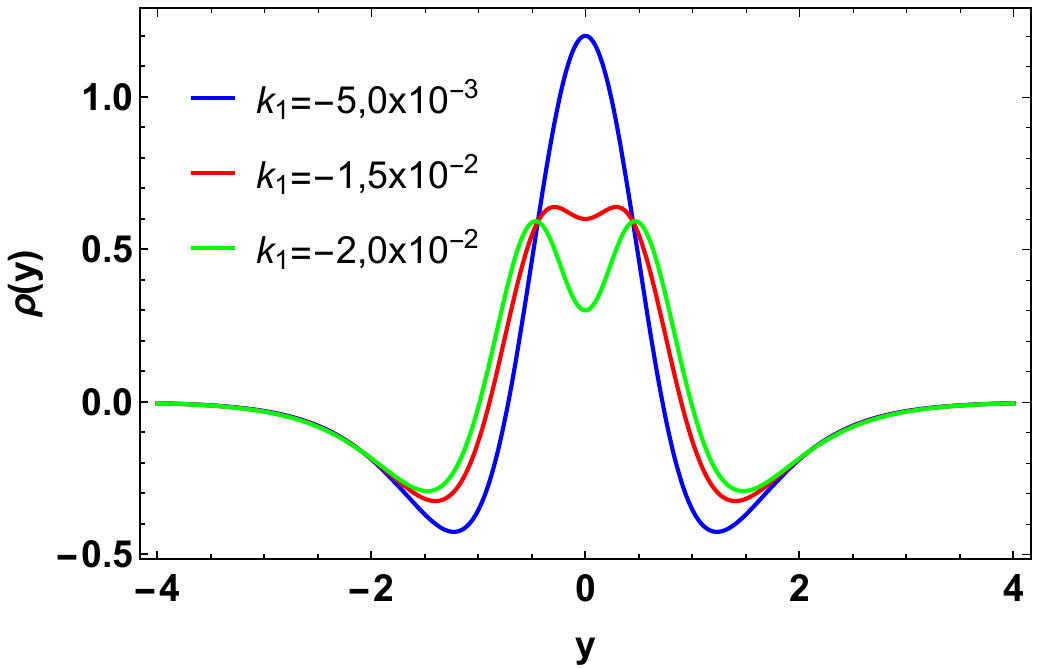}\\
(a)\hspace{7.7cm}(b)\\
\end{tabular}
\end{center}
\vspace{-0.5cm}
\caption{For quadratic gravity. (a) Scalar field. (b) Energy density. 
\label{fig2}}
\end{figure}

It is worth to note that when $k_1=k_2=k_3=0$ (GR case), we obtain the usual kink-like solution, looking like 
\begin{align}
\label{scalarQ}\phi(y)= \sqrt{6p} \arctan\bigg[\tanh\bigg(\frac{\lambda y}{2}\bigg)\bigg].    
\end{align}
The corresponding energy density is given by
\begin{align}
\rho(y)&=\frac{1}{2} \omega\ \sech^{2 \omega }(y) \bigg\{4 \omega ^3 \tanh ^4(y) (10 k_1+2 k_2+k_3)\nonumber\\&+\sech^2(y) \bigg[3-4 \tanh ^2(y) \bigg(\omega ^2 (74 k_1 +22 k_2+17 k_3)\nonumber\\&+4 \omega  (16 k_1+5 k_2+4 k_3)+16 k_1+5 k_2+4 k_3\bigg)\bigg]\nonumber\\&+2 (3 \omega +1) \sech^4(y) (16 k_1+5 k_2+4 k_3)-6 \omega  \tanh ^2(y)\bigg\}.   
\end{align}
As we can see from Fig. (\ref{fig2}), the matter field solution displays a behavior in which the kink-like profile is deformed into a two-kink one. Such a behavior is also suggested by energy density which tends to split as $k_1$ increases. For this analysis, one adopts the relation $k_2=-k_1$ and $k_3=k_1$. It is worth mentioning that two-kink solutions can also appear in the context of GR by considering a multi-field theory or noncanonical dynamics. In our case, the scalar field profile as well as the energy density are influenced mainly by the gravitational effects, i.e., the brane splitting that happens has a pure geometrical origin. 

The next step consists in studying the cubic gravity.

\subsection{Cubic gravity}
As a next step, let us now consider a cubic gravity by  choosing the following form of our function $f(R,Q,P)$:
\begin{align}
 f(R,Q,P)=R+k_1\, R^3+k_2\, R\, Q + k_3\, R\, P.   
\end{align}
In this case, we have terms involving mixing between the Ricci scalar and quadratic invariants $Q,P$. The corresponding matter field equation looks like
\begin{align}
\label{scalarC}\phi^{\prime 2}(y)&=\frac{1}{2}\omega\  \sech^2(y) \bigg\{6 \omega ^2 (15 k_1 (\omega  (32-5 \omega )+16)\nonumber\\&+\omega  (-5 \omega  (2 k_2+k_3)+82 k_2+56 k_3)+41 k_2+28 k_3)\nonumber\\&+8 \omega  \sech^2(y)\bigg[-2 (5 \omega +2) \sech^2(y) (3 k_1 (\omega -20) (5 \omega +2)+k_2 (2 (\omega -25) \omega -25)\nonumber\\&+k_3 ((\omega -34) \omega -20))+12 k_1 (5 \omega +2) (\omega  (5 \omega -66)-16)\nonumber\\&+k_2 (\omega  (8 \omega  (5 \omega -82)-467)-80)+4 k_3 (\omega  (\omega  (5 \omega -112)-85)-16)\bigg]+3\bigg\}.
\end{align}
Like the previous case, due to complexity of above equation, we can solve it only numerically. Additionally, the energy density for our model reads
\begin{align}
\rho(y)&=\frac{1}{2} \omega\  \sech^{2 \omega }(y) \bigg\{80 \omega ^5 (10 k_1-2 k_2-k_3)+\sech^2(y) \bigg[8 \omega  \sech^2(y)\nonumber\\& \bigg(-(5 \omega +2) \sech^2(y) (2 k_1 (\omega -20) (2 \omega +3) (5 \omega +2)-k_2 (\omega  (4 \omega  (\omega +17)+135)+50)\nonumber\\&-2 k_3 (\omega  (\omega  (\omega +23)+48)+20))+12 k_1 (\omega +1) (5 \omega +2)  (\omega  (5 \omega -66)-16)\nonumber\\&-k_2 (\omega  (\omega  (12 \omega  (5 \omega +58)+1139)+547)+80)-2 k_3 (\omega  (\omega  (3 \omega  (5 \omega +78)+398)+202)\nonumber\\&+32)\bigg)+2 \omega  (8 \omega  (15 k_1 (\omega  (\omega  (59-10 \omega )+64)+16)+k_2 (2 \omega  (3 \omega  (5 \omega +29)+82)+41)\nonumber\\&+k_3 (\omega  (3 \omega  (5 \omega +39)+112)+28))+3)+3\bigg]-6 \omega \bigg\}
\end{align}

\begin{figure}[ht!]
\begin{center}
\begin{tabular}{ccc}
\includegraphics[scale=0.45]{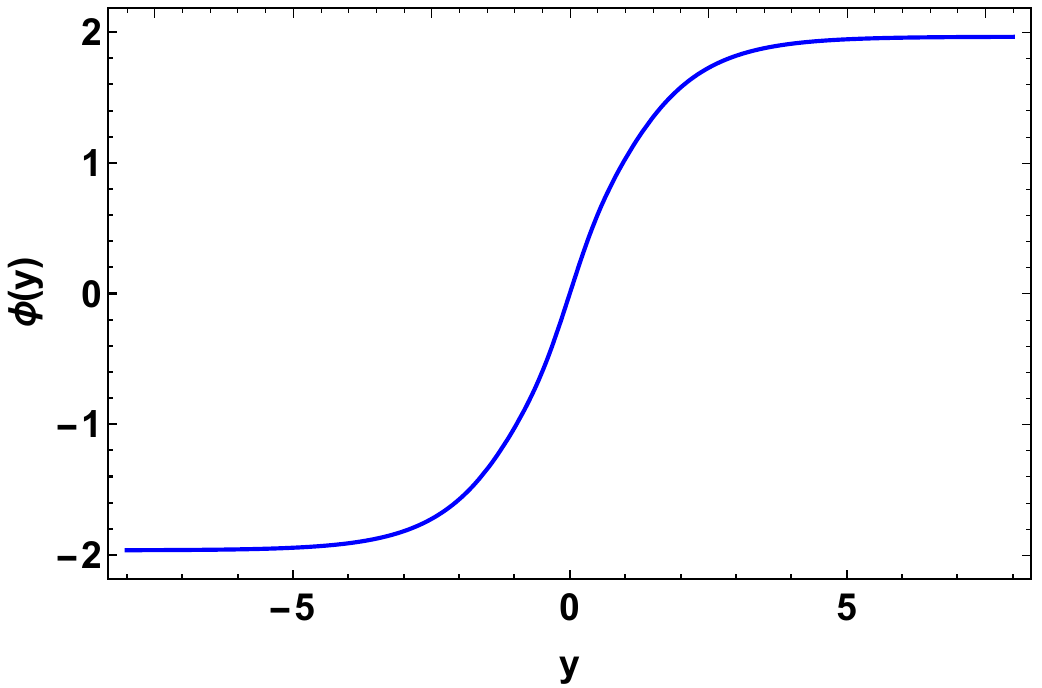} 
\includegraphics[scale=0.45]{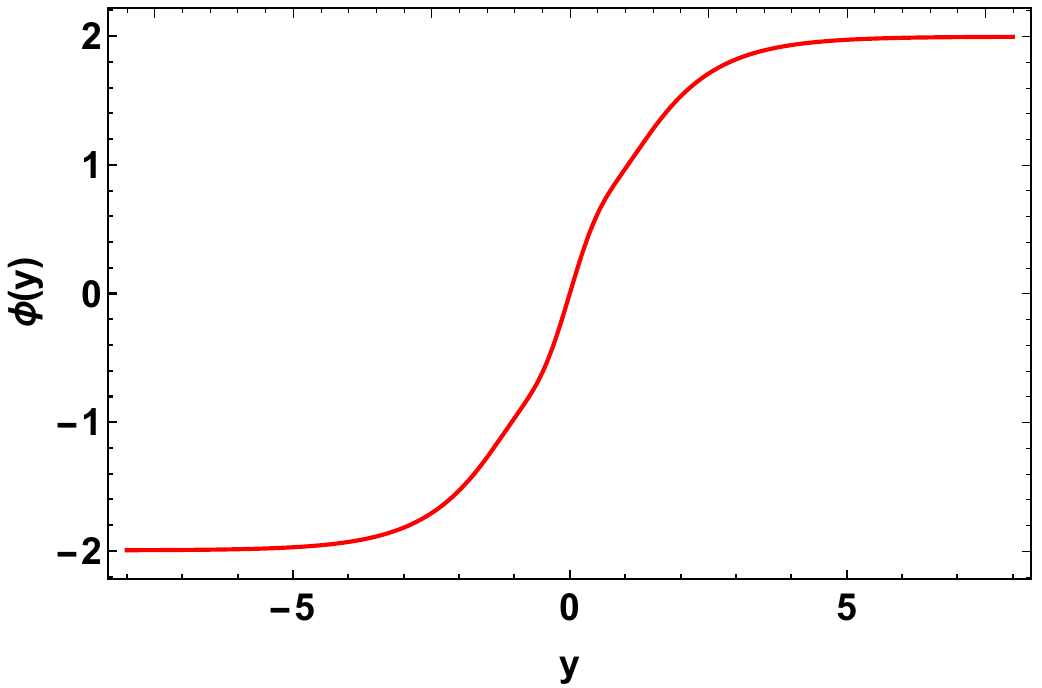}\\
(a)\hspace{6.7cm}(b)\\
\includegraphics[scale=0.45]{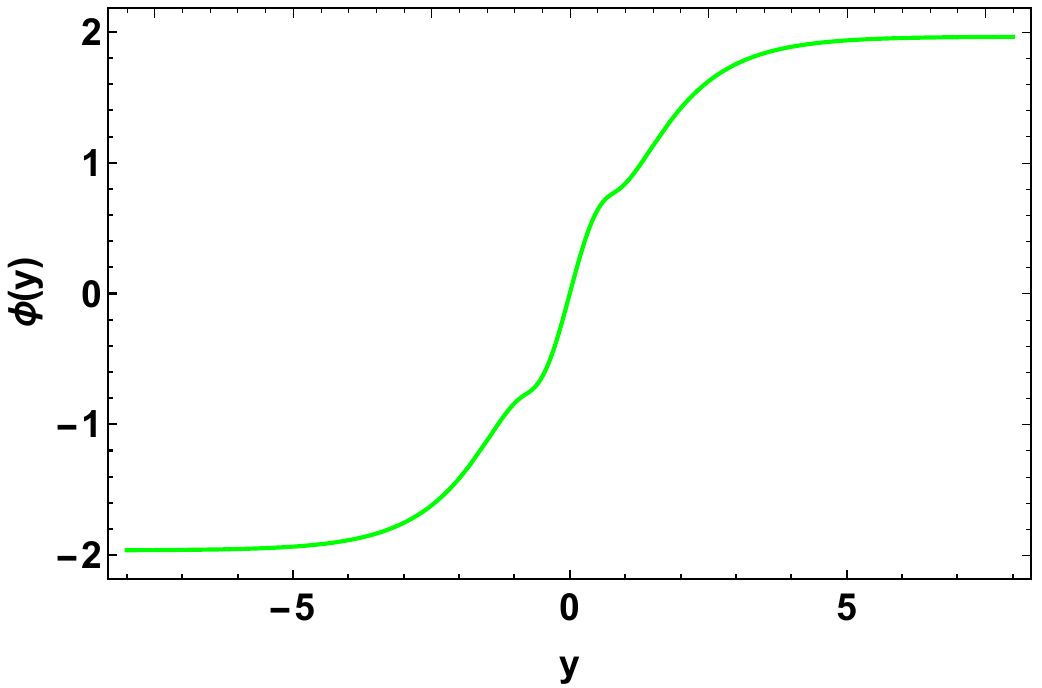} 
\includegraphics[scale=0.45]{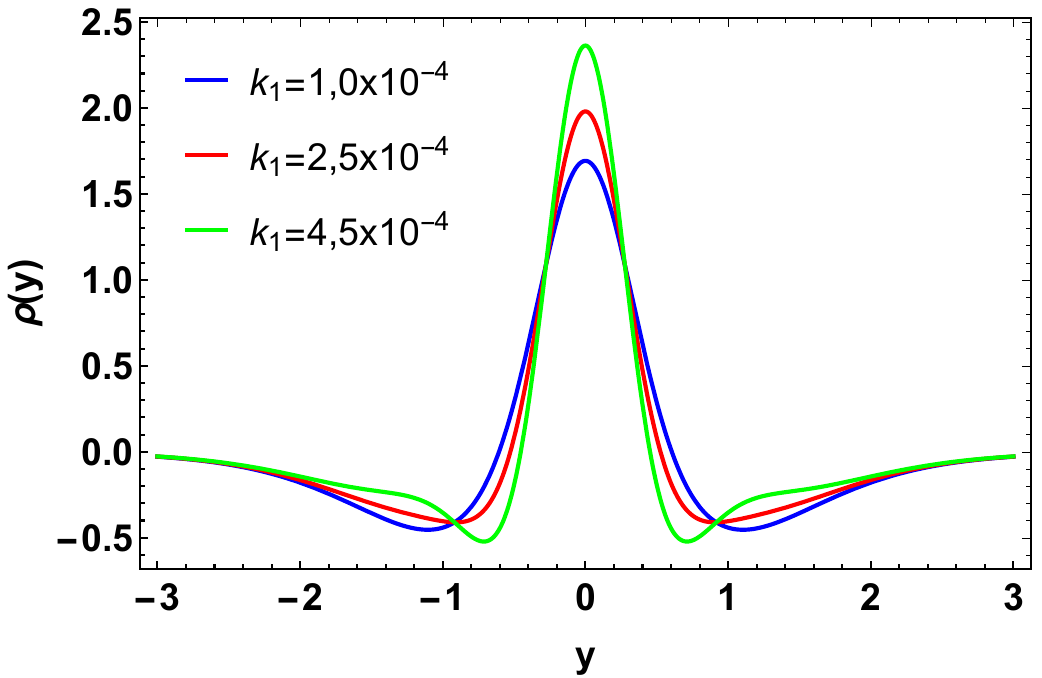}\\
(c)\hspace{6.7cm}(d)\\
\end{tabular}
\end{center}
\vspace{-0.5cm}
\caption{For cubic gravity. (a) Scalar field with $k_1=1,0 \cdot 10^{-4}$. (b) Scalar field with $k_1=2,5 \cdot 10^{-4}$. (c) Scalar field with $k_1=4,5 \cdot 10^{-4}$. (d) Energy density. 
\label{fig3}}
\end{figure}

In Fig. (\ref{fig3}), we plotted the scalar field solution and the brane energy density. Unlike quadratic gravity, the scalar field solution tends to a 3-kink when modifying the parameter $k_1$. In this case, we consider the relation between the parameters as being $k_2=k_1$ and $k_3=k_1$. In flat space-time, a 3-kink structure can be obtained by dealing with two scalar fields \cite{Bazeia:2019vld}. Again, in our conjecture, the emergence of the 3-kink solution is provoked 
mainly by modified gravity.

Once we have discussed the cubic case let us now turn to the quartic gravity.

\subsection{Quartic gravity}
In this scenario, we will take three examples of the function $f(R,Q,P)$. The first one we choose is
\begin{align}\label{gq1}
 f(R,Q,P)=R+k_1\, R^4+k_2\, Q^2 + k_3\,  P^2.  
\end{align}
The equation of motion for this case reads
\begin{align}\label{sfeqq1}
\phi^{\prime 2}(y)&=\frac{3}{2} \omega\,  \sech^2(y) \bigg\{16 \omega ^2 \bigg[4 \omega ^2 \tanh ^6(y) \bigg(5 \omega ^2 (100 k_1+4 k_2+k_3)\nonumber\\&-4 \omega  (400 k_1+19 k_2+6 k_3)-2 (400 k_1+19 k_2+6 k_3)\bigg)\nonumber\\&+\sech^6(y)\, (256 k_1 (3 \omega +1)+60 k_2 \omega +25 k_2+32 k_3 \omega +16 k_3)\nonumber\\&-2 \tanh ^2(y)\, \sech^4(y) (32 k_1 (\omega  (155 \omega +116)+24)\nonumber\\&+2 \omega  (137 k_2 \omega +125 k_2+50 k_3 \omega +56 k_3)+75 k_2+48 k_3)\nonumber\\&+4 \omega\,  \tanh ^4(y)\, \sech^2(y) (80 k_1 (\omega  (60 \omega +61)+16)\nonumber\\&+\omega  (228 k_2 \omega +251 k_2+72 k_3 \omega +86 k_3)+80 k_2+32 k_3)\bigg]+1\bigg\}    
\end{align}

\begin{figure}[ht!]
\begin{center}
\begin{tabular}{ccc}
\includegraphics[scale=0.45]{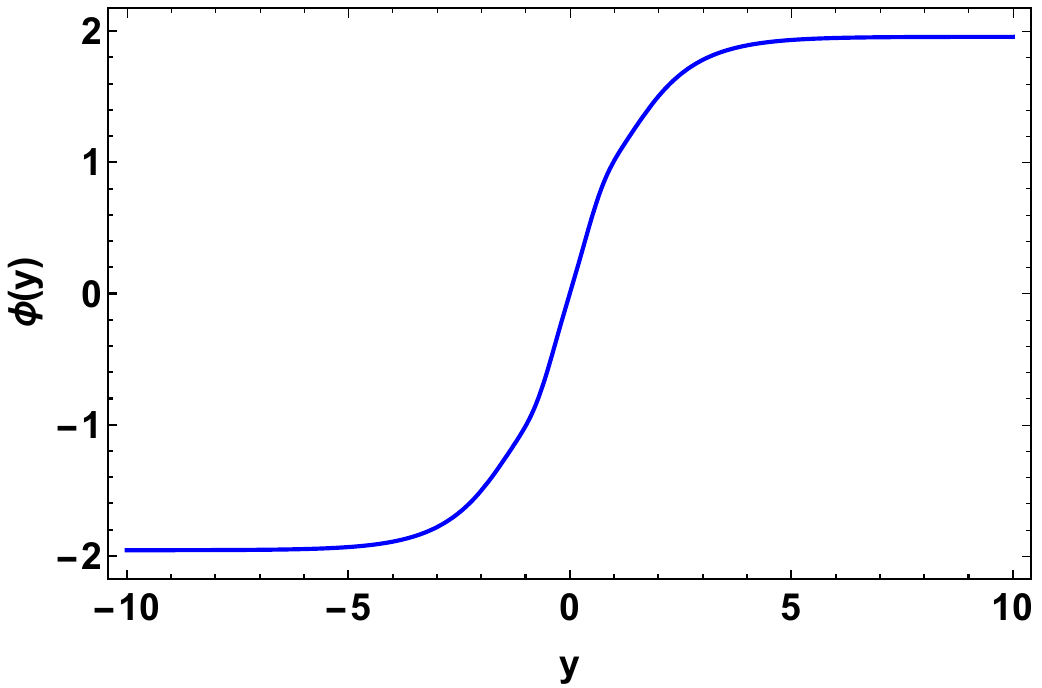} 
\includegraphics[scale=0.45]{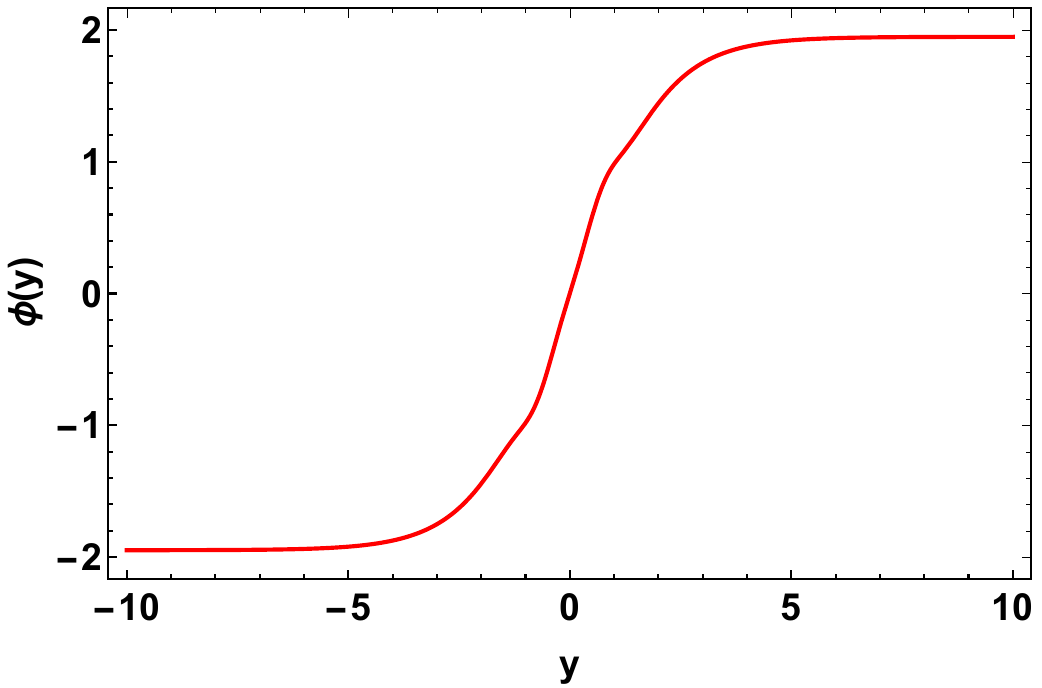}\\
(a)\hspace{6.7cm}(b)\\
\includegraphics[scale=0.45]{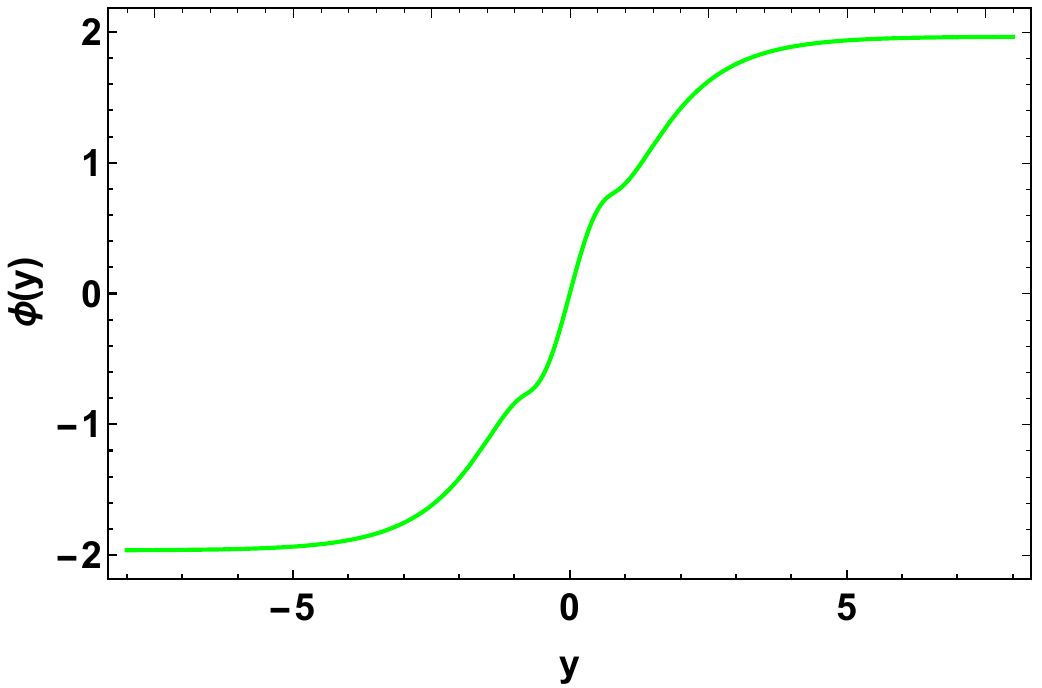} 
\includegraphics[scale=0.45]{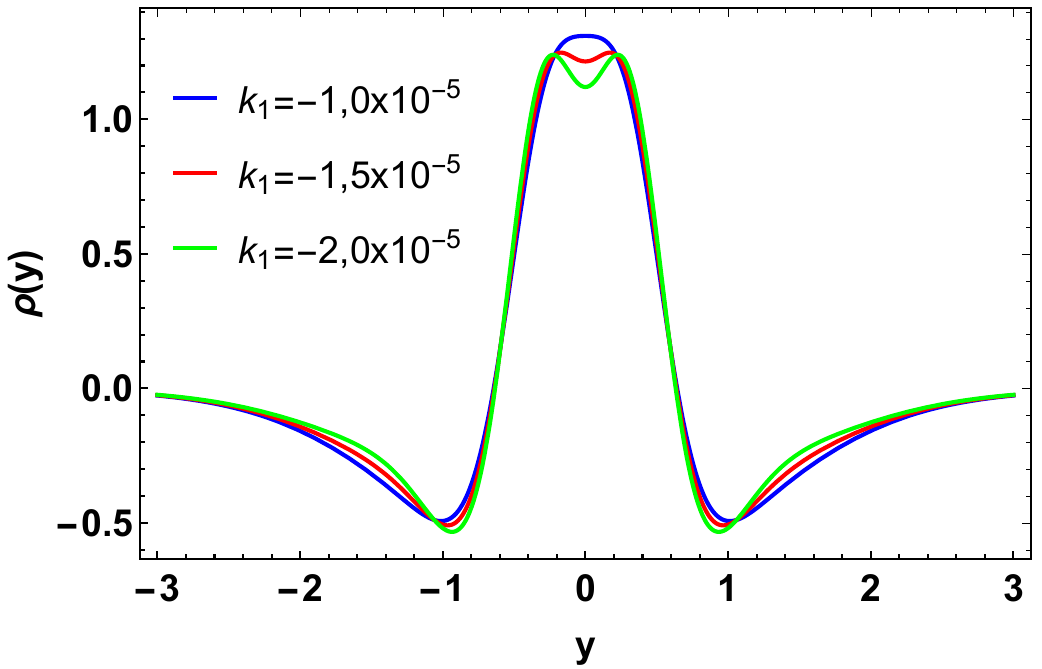}\\
(c)\hspace{6.7cm}(d)\\
\end{tabular}
\end{center}
\vspace{-0.5cm}
\caption{For quartic gravity  (\ref{gq1}). (a) Scalar field with $k_1=-1,0 \cdot 10^{-5}$. (b) Scalar field with $k_1=-1,5 \cdot 10^{-5}$. (c) Scalar field with $k_1=-2,0 \cdot 10^{-5}$. (d) Energy density. 
\label{fig4}}
\end{figure}

The complete expression for the energy density is extremely cumbersome, so we will only present its qualitative behavior through Fig. (\ref{fig4}). Thus, by solving numerically the Eq.(\ref{sfeqq1}), we see the emergence of a 3-kink-like structure also in quartic gravity. In this case, the energy density presents a small split. The relation between the parameters is $k_2=-2k_1$ and $k_3=k_1$. 

The second example of $f(R,Q,P)$ is given by
\begin{align}\label{gq2}
 f(R,Q,P)=R+k_1\, R^4+k_2\, Q\, P.   
\end{align}
In this case the equations of motion yield
\begin{align}\label{sfeqq2}
\phi^{\prime 2}(y)&=\frac{3}{2} \omega\,  \sech^2(y) \bigg\{16 \omega ^2 \bigg[2 \omega ^2 \tanh ^6(y) \bigg(200 k_1 (\omega  (5 \omega -16)-8)\nonumber\\&+k_2 \bigg(20 \omega ^2-86 \omega -43\bigg)\bigg)+4 \sech^6(y) (64 k_1 (3 \omega +1)\nonumber\\&+k_2 (11 \omega +5))-\tanh ^2(y)\, \sech^4(y) (64 k_1 (\omega  (155 \omega +116)+24)\nonumber\\&+k_2 (\omega  (337 \omega +340)+120))+\omega  \tanh ^4(y)\, \sech^2(y) (320 k_1 (\omega  (60 \omega +61)+16)\nonumber\\&+k_2 (\omega  (516 \omega +595)+208))\bigg]+1\bigg\}.
\end{align}

\begin{figure}[ht!]
\begin{center}
\begin{tabular}{ccc}
\includegraphics[scale=0.45]{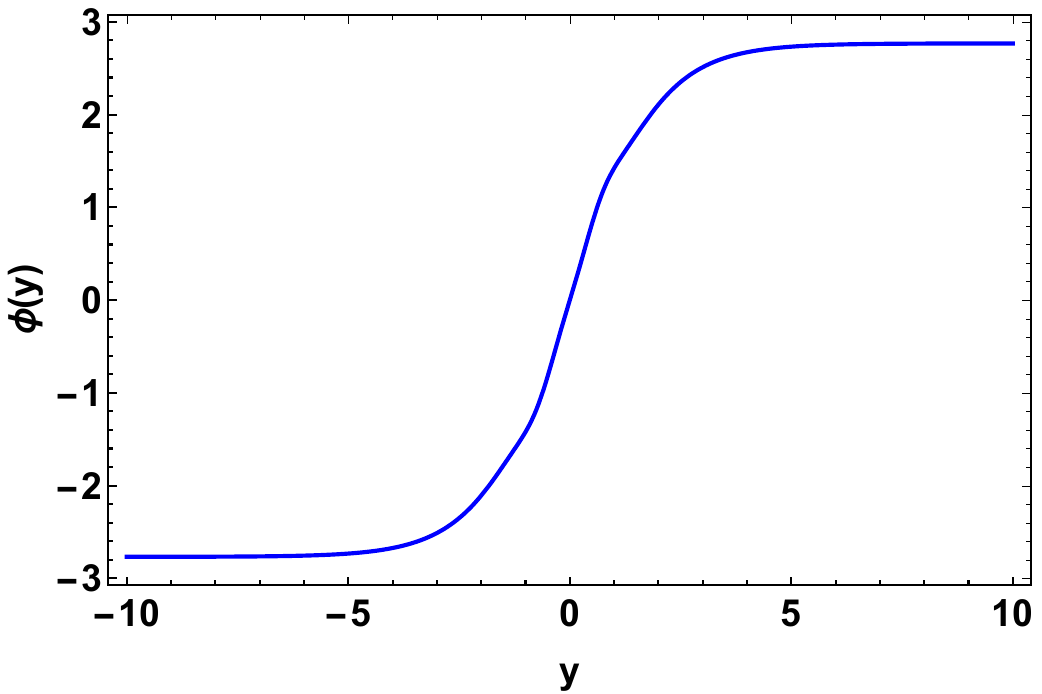} 
\includegraphics[scale=0.45]{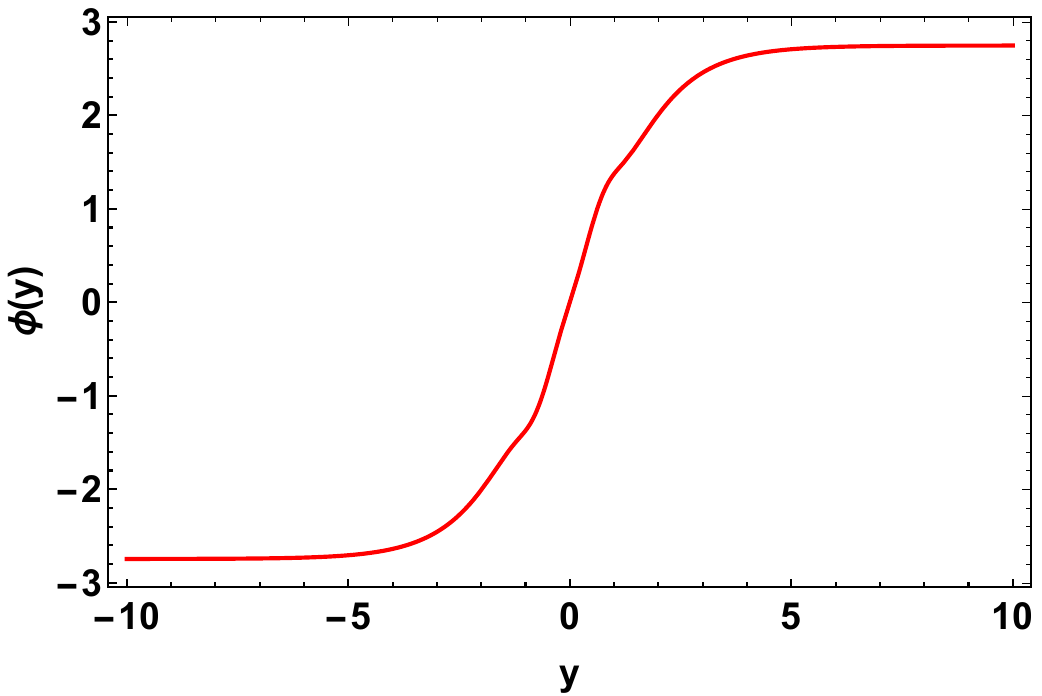}\\
(a)\hspace{6.7cm}(b)\\
\includegraphics[scale=0.45]{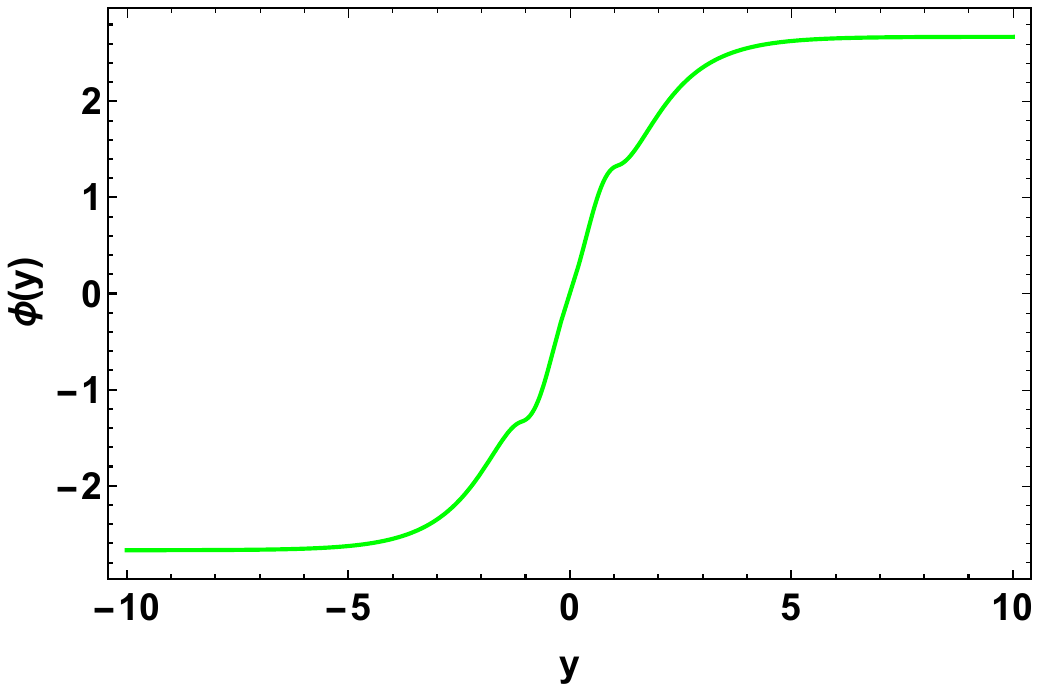} 
\includegraphics[scale=0.45]{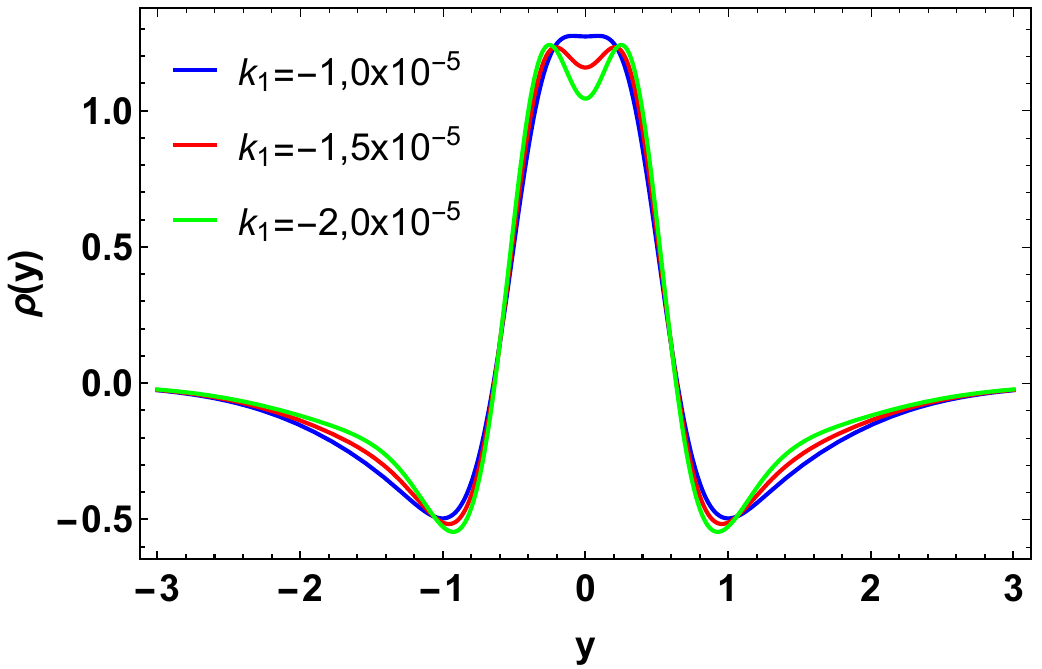}\\
(c)\hspace{6.7cm}(d)\\
\end{tabular}
\end{center}
\vspace{-0.5cm}
\caption{For quartic gravity  (\ref{gq2}). (a) Scalar field with $k_1=-1,0 \cdot 10^{-5}$. (b) Scalar field with $k_1=-1,5 \cdot 10^{-5}$. (c) Scalar field with $k_1=-2,0 \cdot 10^{-5}$. (d) Energy density. 
\label{fig5}}
\end{figure}

The qualitative behavior of the energy density is given by Fig.(\ref{fig5}). We realize that a 3-kink-like solution also emerges and the energy density has a deeper split than the previous case as the parameters $k_1$ increases, where $k_2=k_1$.

 Our last example of the function $f(R,Q,P)$ is given by
\begin{align}\label{gq3}
 f(R,Q,P)=R+(k_1\, R^2+k_2\, Q + k_3\, P)^2, 
\end{align}
 and our equations of motion in this case look like
\begin{align}\label{sfeqq3}
\phi^{\prime 2}(y)&=-\frac{3}{2}  \omega\,  \sech^2(y) \bigg\{16 \omega ^2 \bigg[4 \omega ^2 \tanh ^6(y) (10k_1+2 k_2+k_3) \bigg(5 \omega ^2 (10 k_1\nonumber\\&+2 k_2+k_3)-2 \omega  (80 k_1+19 k_2+12 k_3)-80 k_1-19 k_2-12 k_3\bigg)\nonumber\\&-2 \tanh ^2(y) \sech^4(y) \bigg(\omega ^2 (10 k_1+2 k_2+k_3) (496 k_1+137 k_2+100 k_3)\nonumber\\&+2 \omega  (116 k_1+25 k_2+14 k_3) (16 k_1+5 k_2+4 k_3)+3 (16 k_1+5 k_2+4 k_3)^2\bigg)\nonumber\\&+\sech^6(y) (16 k_1+5 k_2+4 k_3) (16 k_1 (3 \omega +1)+12 k_2 \omega +5 k_2\nonumber\\&+8 k_3 \omega +4 k_3)+2 \omega  \tanh ^4(y)\, \sech^2(y) (10 k_1+2 k_2+k_3) (16 k_1 (\omega  (60 \omega +61)+16)\nonumber\\&+k_2 (\omega  (228 \omega +251)+80)+4 k_3 (\omega  (36 \omega +43)+16))\bigg]-1\bigg\}
\end{align}

\begin{figure}[ht!]
\begin{center}
\begin{tabular}{ccc}
\includegraphics[scale=0.45]{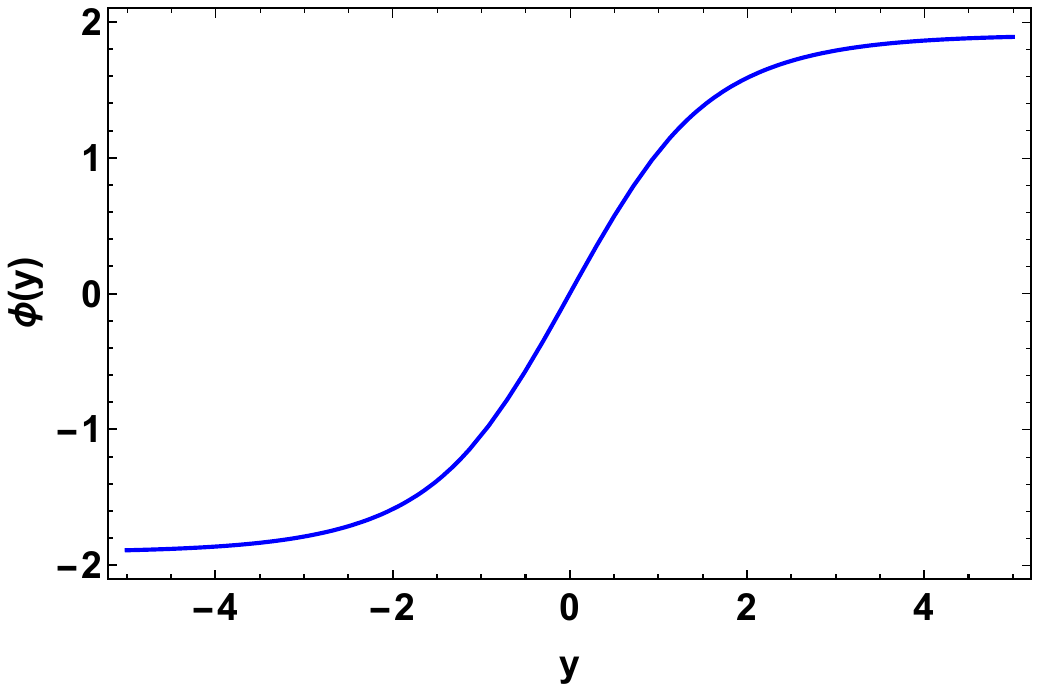} 
\includegraphics[scale=0.45]{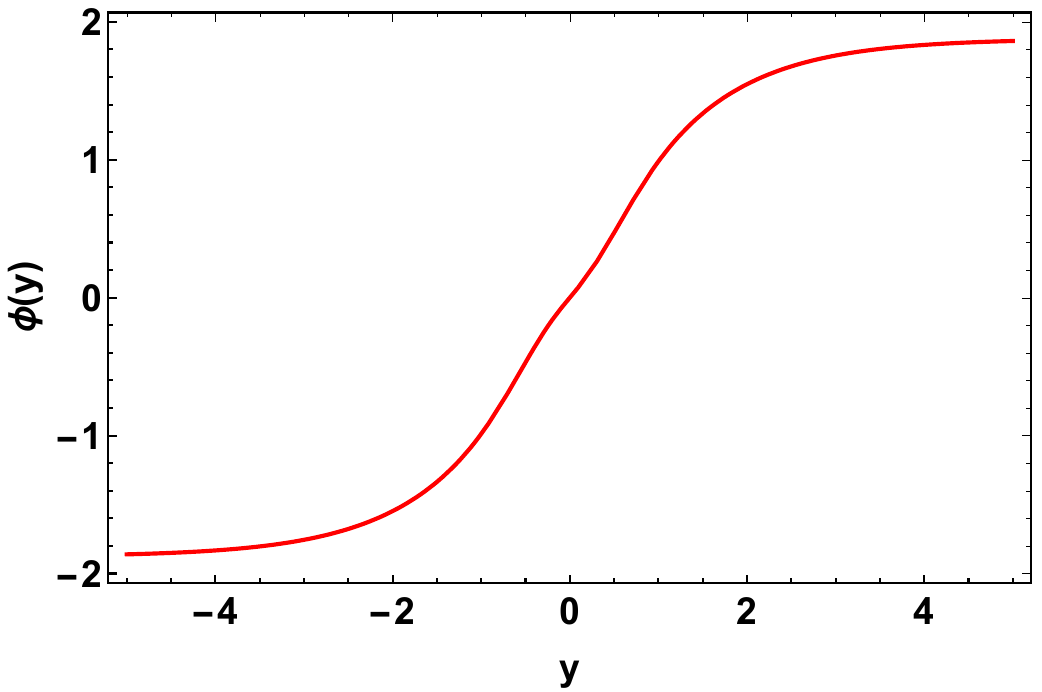}\\
(a)\hspace{6.7cm}(b)\\
\includegraphics[scale=0.45]{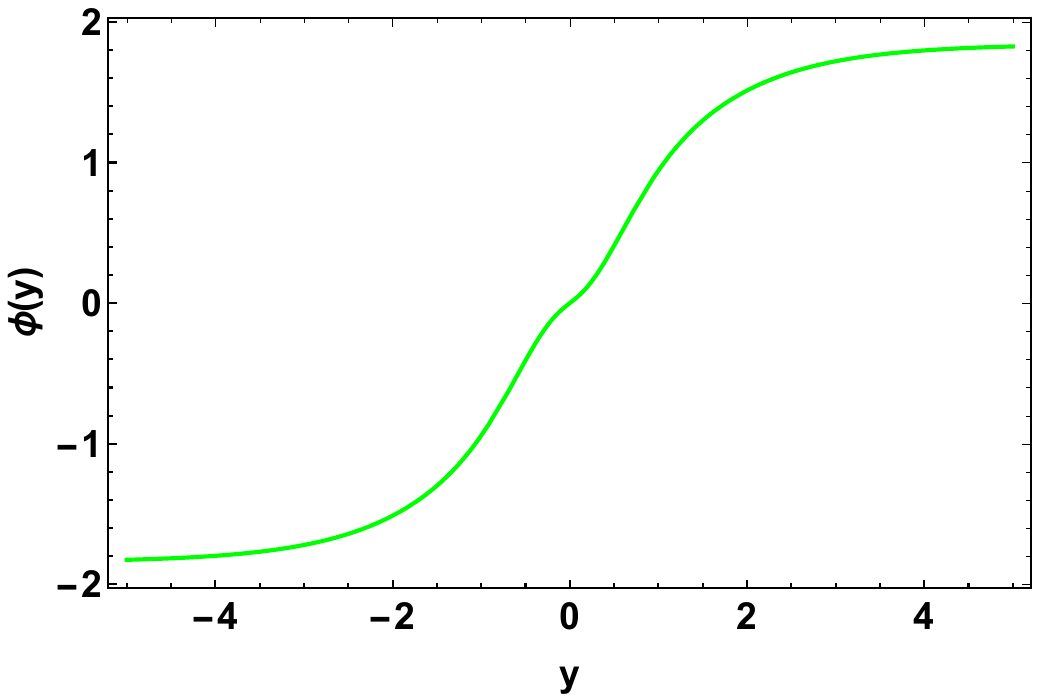} 
\includegraphics[scale=0.45]{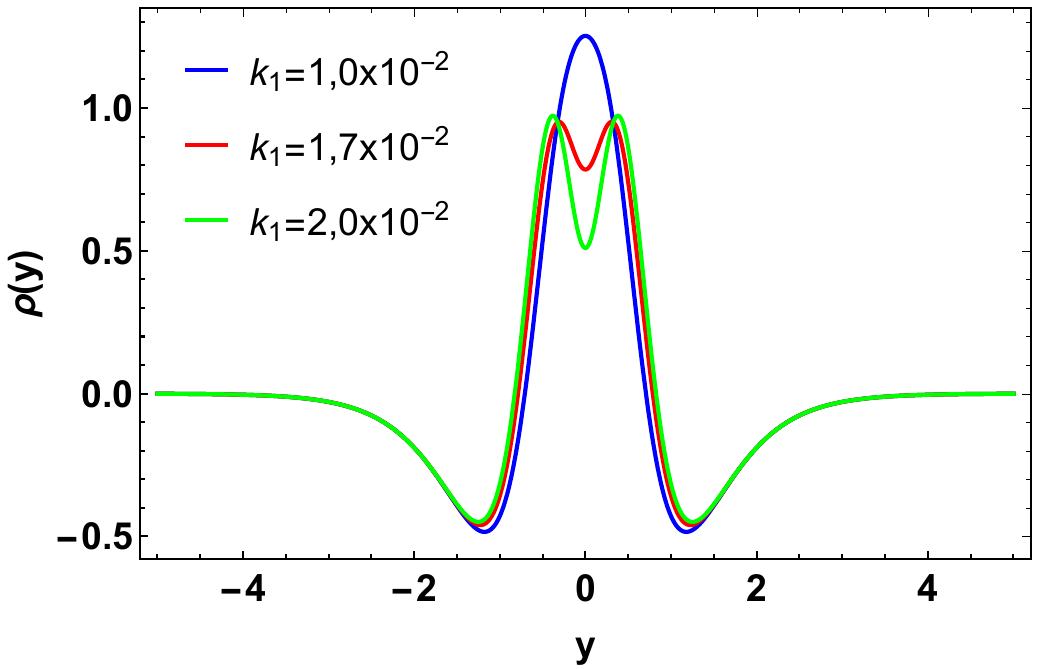}\\
(c)\hspace{6.7cm}(d)\\
\end{tabular}
\end{center}
\vspace{-0.5cm}
\caption{For quartic gravity  (\ref{gq3}). (a) Scalar field with $k_1=1,0 \cdot 10^{-2}$. (b) Scalar field with $k_1=1,7 \cdot 10^{-2}$. (c) Scalar field with $k_1=2,0 \cdot 10^{-2}$. (d) Energy density. 
\label{fig6}}
\end{figure}

These results can be presented graphically in Fig.(\ref{fig6}). Unlike the previous case, one notes that now the scalar field possesses a 2-kink-like profile. Again, the energy density presents an even deeper split when we vary the parameter $k_1$, where $k_2=5,5 k_1$ and $k_3=k_1$.

%%%%%%%%%%%%%%%%%%%%%%%%%%%%%%%%%%%%%%%%%%%%%%%%%%%%%%%%%%%%%%%%%%%%%%%%
\section{Dirac fermion localization}\label{s4}

%Besides studying the scalar field solution, we should analyze the consistency of {\bf the} theory. For this purpose, the localization of fermions will be addressed within the braneworld context.

In this section, we will investigate the influence of a deformed scalar field solution on the localization of spin 1/2 fermions. Let us then assume the following 5D Dirac action with the Yukawa-like coupling with matter field \cite{Liu:2009ve,Melfo:2006hh,Liu:2008pi,Liu:2009dw,Chumbes:2010xg,Almeida:2009jc,Cruz:2011ru}:
\begin{align}
    S_{\frac{1}{2}}=\int d^5x\sqrt{-g}\,(i\,\overline{\Psi}\Gamma^M D_M \Psi-\lambda\phi\,\overline{\Psi}\Psi),
\end{align}
where $\lambda$ is a positive constant that represents the coupling between fermions
and the scalar field. It is noteworthy that the scalar field $\phi$ must be interpreted as a background field. Such a coupling is quite important for the brane scenario to support normalized zero-mode for the fermionic field. Additionally, $\Gamma^M$ represent the Dirac matrices in a curved space-time, where $\Gamma^M=h_a\ ^M\gamma^a$ and $\{\Gamma^M,\Gamma^N\}=2\eta^{MN}$. It is worth mentioning that $\gamma^a$ stands for the gamma matrices in the flat space-time. Finally, $D_M=\partial_M+\Omega_M$ is the covariant derivative, with $\Omega_M$ being defined as follows
\begin{align}
\Omega_M=\frac{1}{4}\omega_M\ ^{ab}\gamma_a\gamma_b,    
\end{align}
where $\omega_M^{ab}$ is the spin connection. By varying the action with respect to $\overline{\Psi}$, the equation of motion reads
\begin{align}\label{fermionseq}
i\,\Gamma^M D_M \Psi-\lambda\phi\Psi=0.    
\end{align}
In order to deal with the fermion localization, it is convenient to consider the conformal coordinate $z$, where $dy=\e^{A(z)}dz$, so that the braneworld metric is recast as
\begin{align}\label{newmetric}
    ds^2=\e^{2A(z)}(\eta_{\mu\nu}dx^\mu dx^\nu+dz^2)
\end{align}
For this new metric, one employs the representation for gamma matrices $\Gamma^M=(\e^{-A}\gamma^\mu,-i\e^{-A}\gamma^4)$. Thus, the nonvanishing spin connection component reads as
\begin{align}
\Omega_\mu=\frac{1}{2}\Dot{A} \gamma_\mu\gamma_4   
\end{align}
Thus, we rewrite the Eq.(\ref{fermionseq}) as
\begin{align}
[i\gamma^\mu \partial_\mu + \gamma^4(\partial_z+2\Dot{A})+\lambda\phi\, \e^{A}]\Psi(x,z) =0
\end{align}
To proceed further, we apply the Kaluza-Klein decomposition with two  chiralities, i.e.,
\begin{align}
\Psi(x,z)=\sum_n[\psi_{L,n}(x)\,\chi_{L,n}(z)+\psi_{R,n}(x)\,\chi_{R,n}(z)]\,\e^{-2A}. 
\end{align}
In the above decomposition, $\psi_{L,n}(x)$ and $\psi_{R,n}(x)$ represent, respectively, the left-handed and right-handed components of the 4D spinor field with mass $m$. Besides, there are two scalars $\chi_{L,n}(z)$ and $\chi_{R,n}(z)$ that depend only on the extra dimension $z$. In addition, one writes the properties of $\psi_{L,n}$ and $\psi_{R,n}$ as
\begin{align}
i\gamma^\mu \partial_\mu\psi_{L,n}=m \psi_{Rn} , \ i\gamma^\mu \partial_\mu\psi_{R,n}=m\psi_{L,n},    
\end{align}
and
\begin{align}
\gamma^4 \psi_{L,n}=-\psi_{L,n}    , \  \gamma^4 \psi_{R,n}=\psi_{R,n}.
\end{align}
Making use of the chiral decomposition as well as the above properties, we obtain the following set of coupled equations
\begin{align}
    &(\partial_z-\lambda\phi\, \e^{A})\chi_{L}=m\chi_{R},\\
    &(\partial_z+\lambda\phi\, \e^{A})\chi_{R}=-m\chi_{L},
\end{align}
which can be rewritten in the Schr\"{o}dinger-like form
\begin{align}
    \label{left}&(-\partial_z^2+V_L(z))\chi_{L}=m^2\chi_{L},\\
   \label{right} &(-\partial_z^2+V_R(z))\chi_{R}=m^2\chi_{R},
\end{align}
where we have defined the effective potentials
\begin{align}
V_L=(\lambda\phi e^{A})^2-\partial_z(\lambda\phi e^{A}),\\
V_R=(\lambda\phi e^{A})^2+\partial_z(\lambda\phi e^{A}). 
\end{align}
Therefore, the solution of Eq. \eqref{left} is the left-handed zero-mode looking like
\begin{align}
\chi_{L}=e^{-\lambda\int dz \,\phi(z) },  
\end{align}
while the solution of Eq. \eqref{right} is the right-handed zero-mode of the form
\begin{align}
\chi_{R}=e^{\lambda\int dz\, \phi(z)}.    
\end{align}

\begin{figure}[ht!]
\begin{center}
\begin{tabular}{ccc}
\includegraphics[scale=0.45]{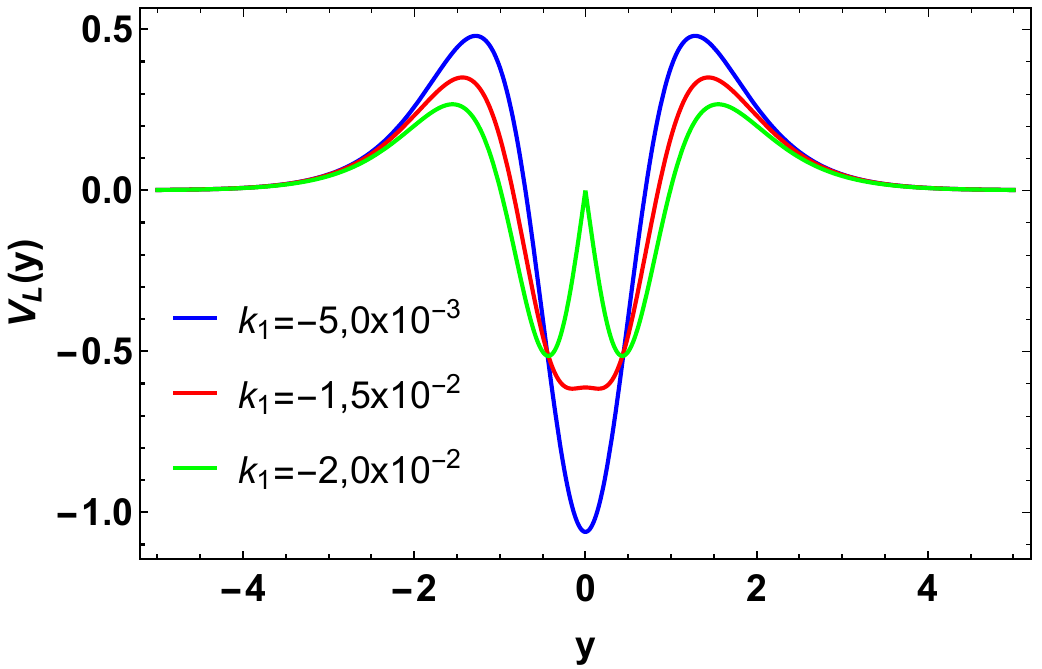} 
\includegraphics[scale=0.45]{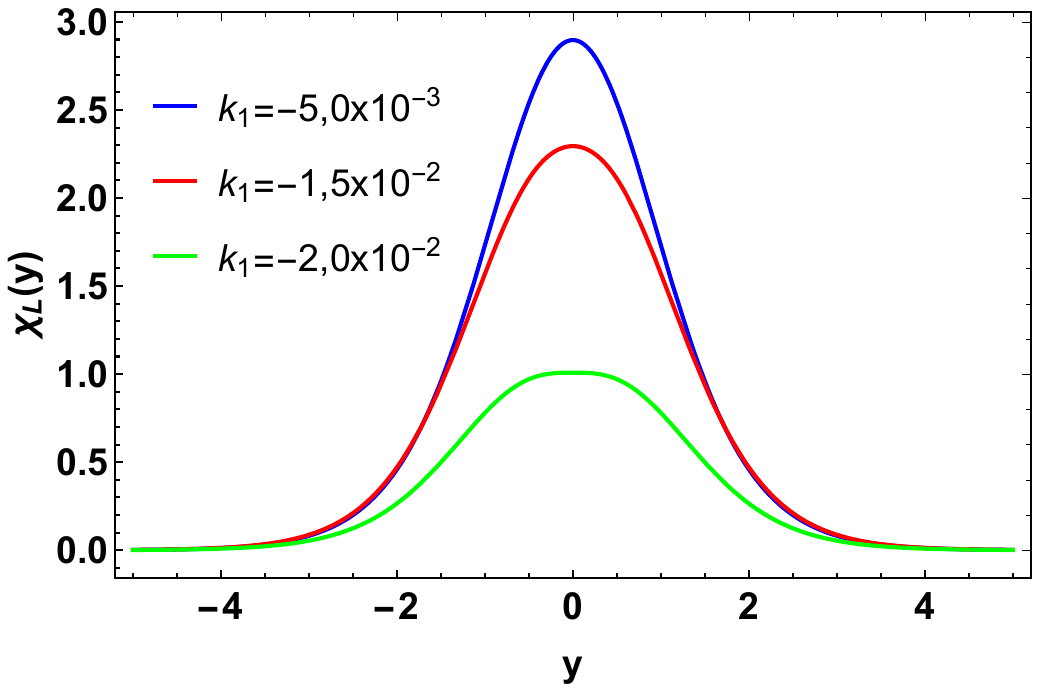}\\
(a)\hspace{6.7cm}(b)\\
\includegraphics[scale=0.45]{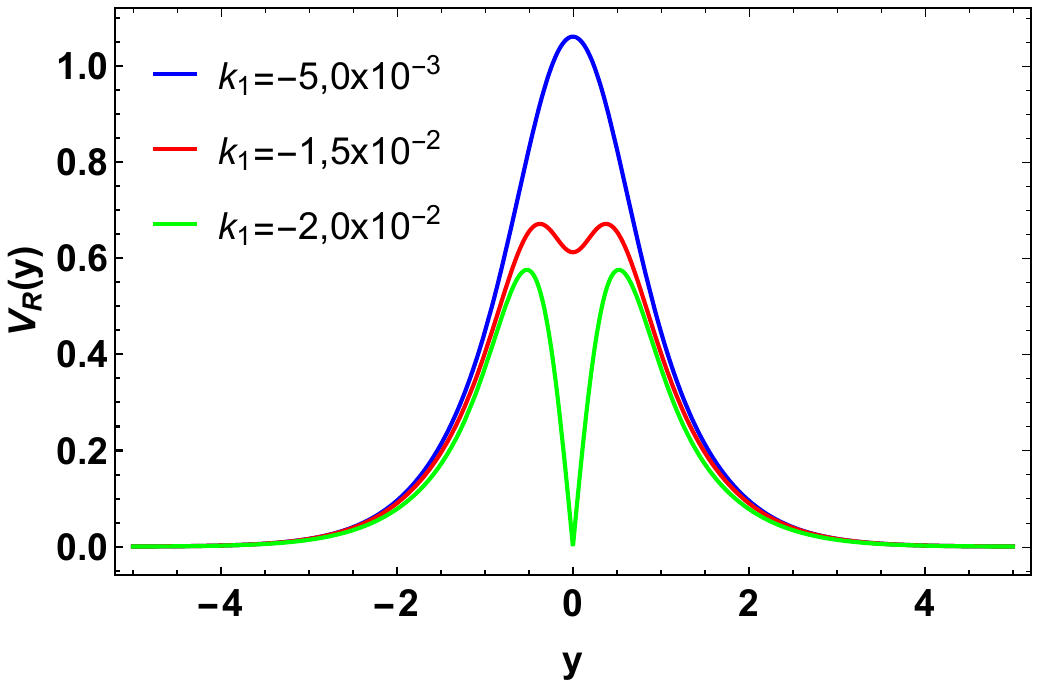} 
\includegraphics[scale=0.45]{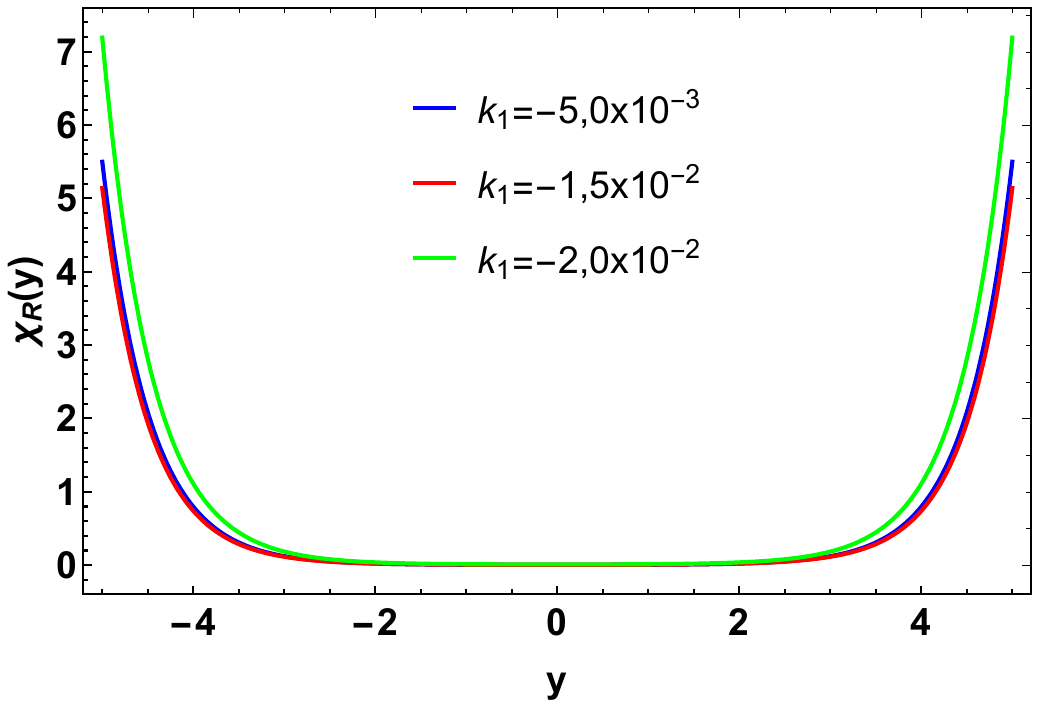}\\
(c)\hspace{6.7cm}(d)\\
\end{tabular}
\end{center}
\vspace{-0.5cm}
\caption{For the scalar field profile \eqref{scalarQ}. (a) left-handed potential. (b) left-handed zero-mode. (c) right-handed potential. (d) right-handed zero-mode. 
\label{fig7}}
\end{figure}

\begin{figure}[ht!]
\begin{center}
\begin{tabular}{ccc}
\includegraphics[scale=0.45]{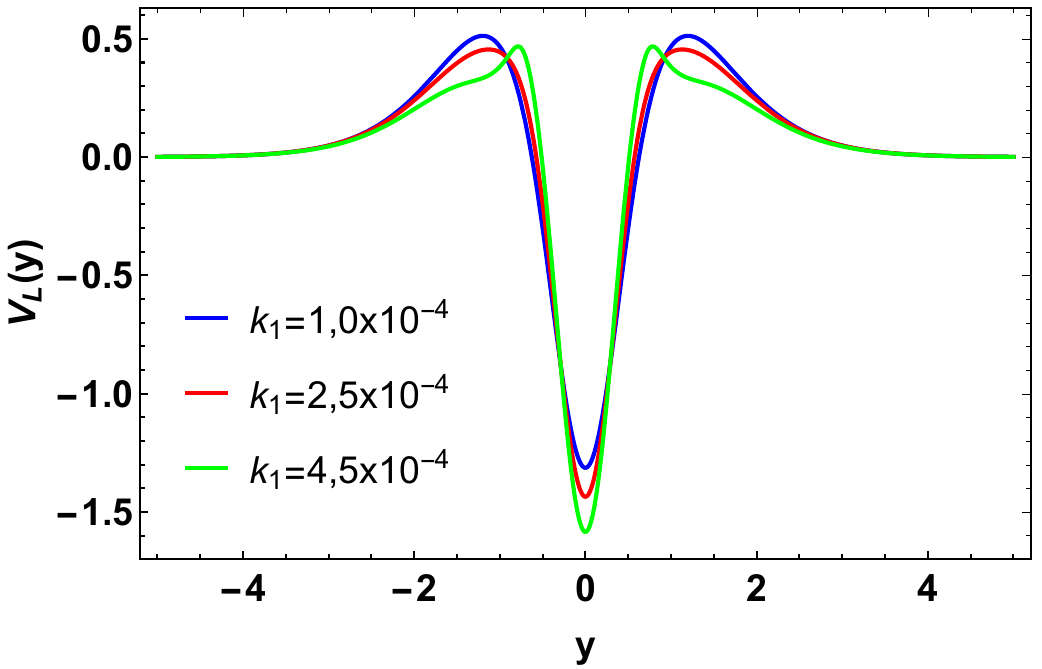} 
\includegraphics[scale=0.45]{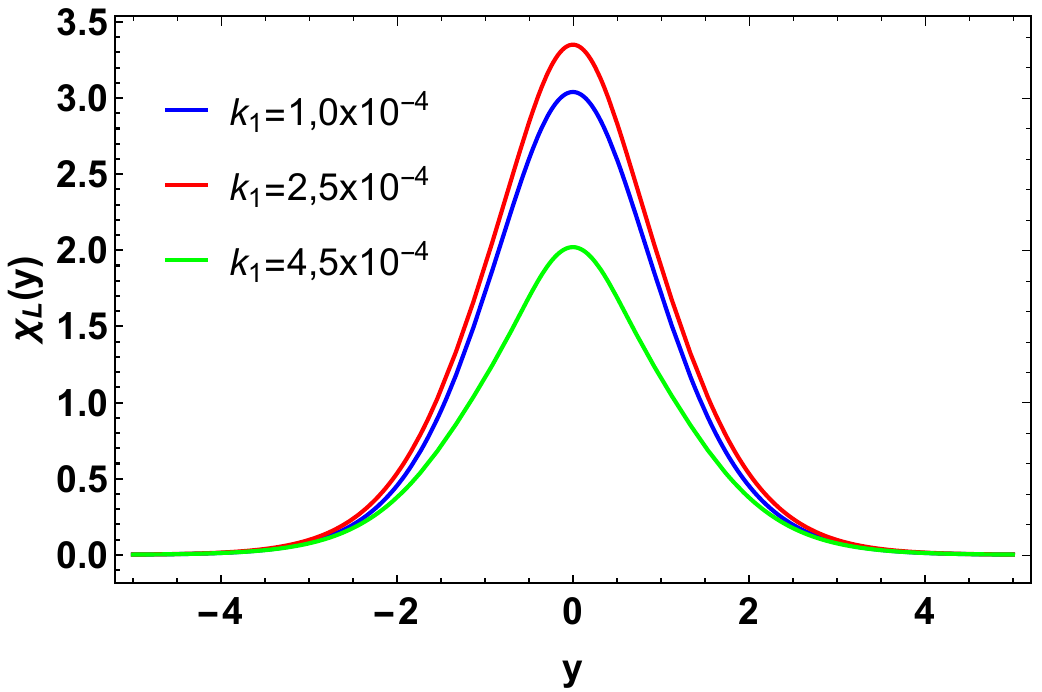}\\
(a)\hspace{6.7cm}(b)\\
\end{tabular}
\end{center}
\vspace{-0.5cm}
\caption{For the scalar field profile given by Eq. \eqref{scalarC}. (a) left-handed potential. (b) left-handed zero-mode. 
\label{fig8}}
\end{figure}

\begin{figure}[ht!]
\begin{center}
\begin{tabular}{ccc}
\includegraphics[scale=0.45]{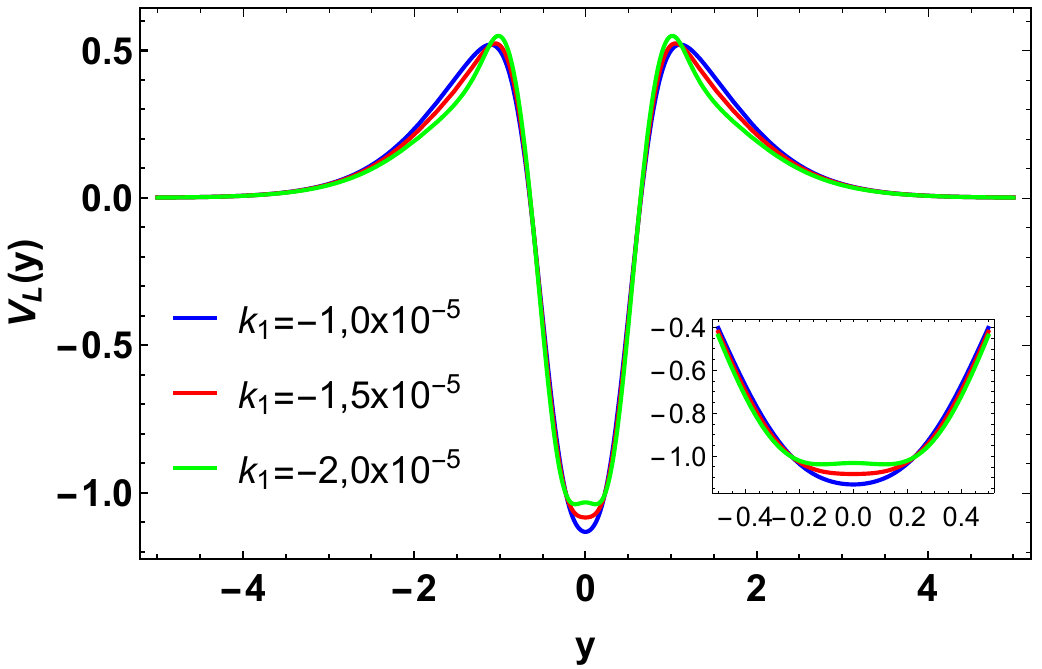} 
\includegraphics[scale=0.45]{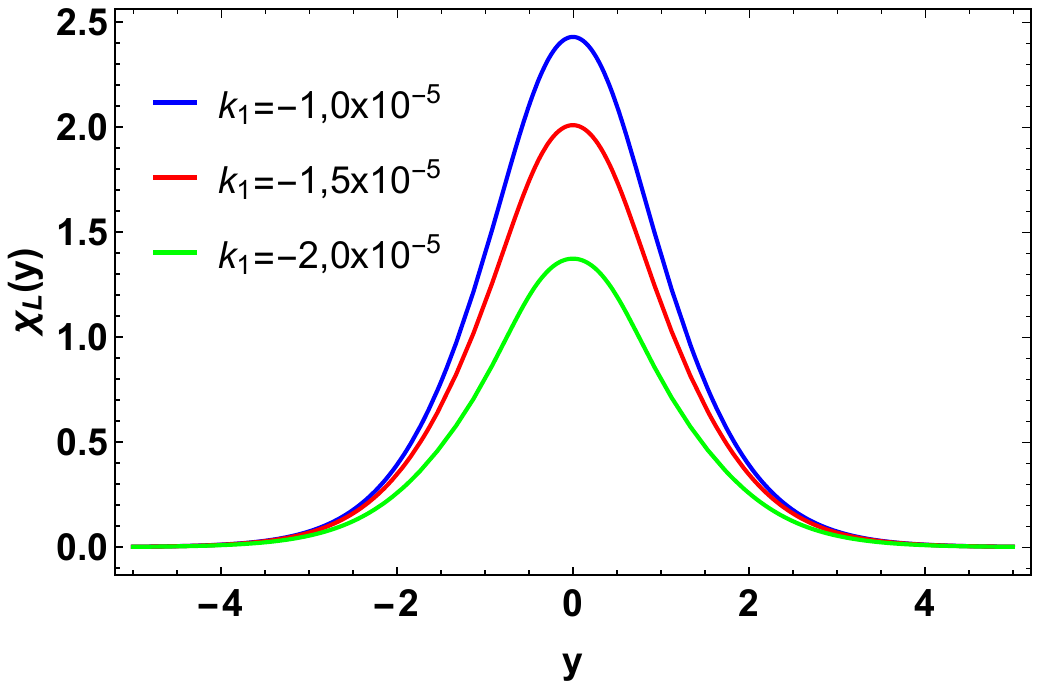}\\
(a)\hspace{6.7cm}(b)\\
\end{tabular}
\end{center}
\vspace{-0.5cm}
\caption{For the scalar field profile given by Eq. (\ref{sfeqq1}). (a) left-handed potential. (b) left-handed zero-mode. 
\label{fig9}}
\end{figure}

\begin{figure}[ht!]
\begin{center}
\begin{tabular}{ccc}
\includegraphics[scale=0.45]{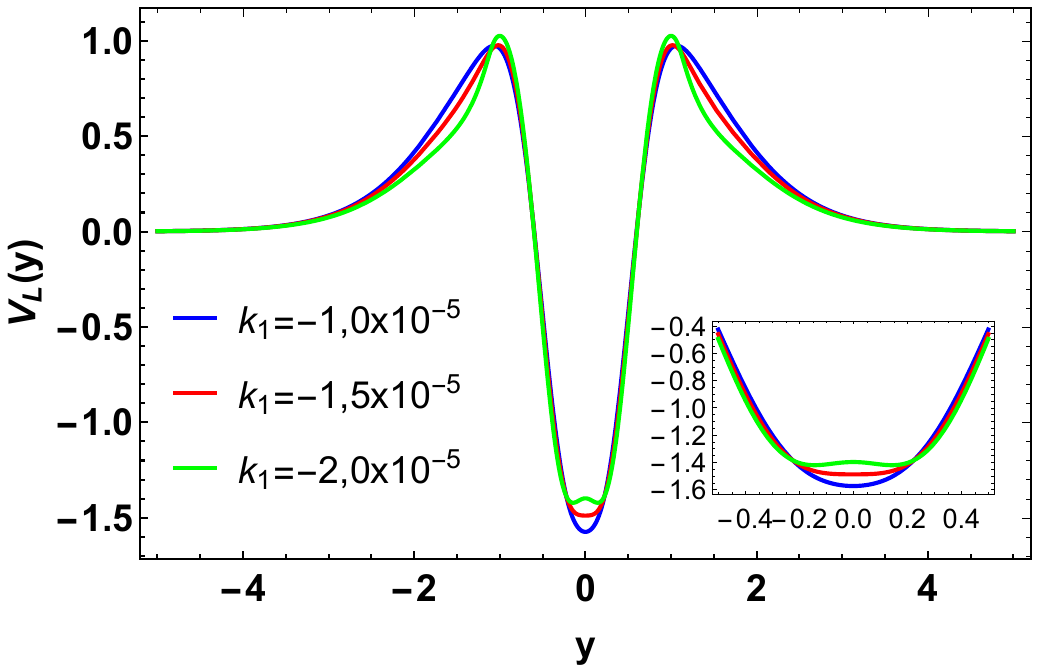} 
\includegraphics[scale=0.45]{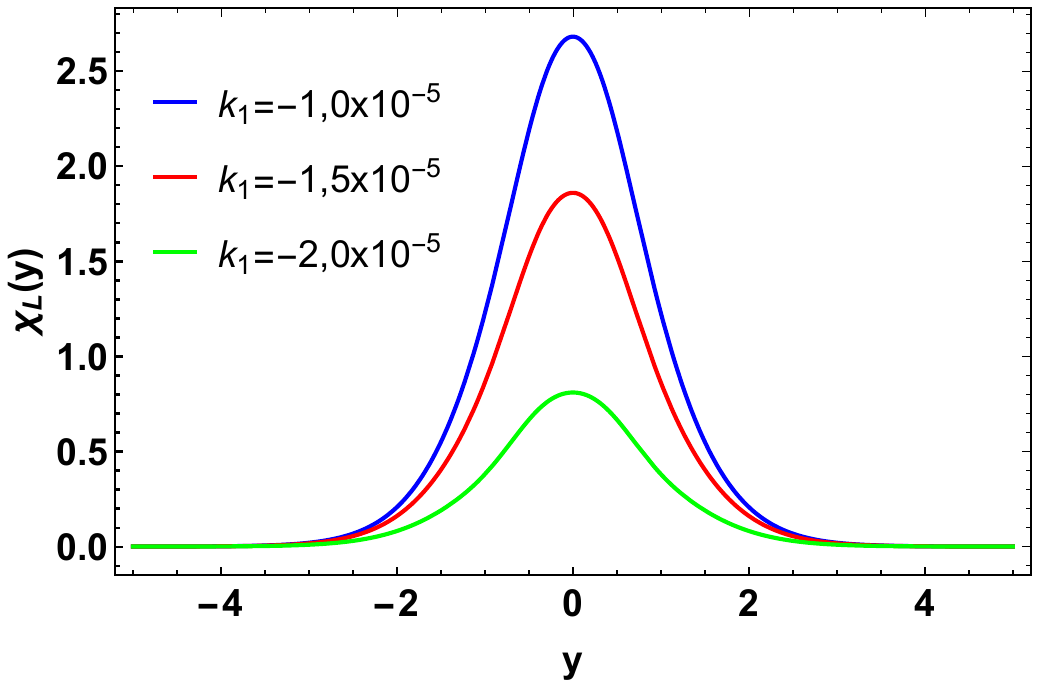}\\
(a)\hspace{6.7cm}(b)\\
\end{tabular}
\end{center}
\vspace{-0.5cm}
\caption{For  the scalar field profile given by Eq. (\ref{sfeqq2}). (a) left-handed potential. (b) left-handed zero-mode.   
\label{fig10}}
\end{figure}

In the Fig. (\ref{fig7}), we depict the behavior of effective potentials and zero-mode for both chiralities and the scalar field profile Eq.\eqref{scalarQ} corresponding to the solution found in the quadratic gravity. As we can observe only the left-handed zero mode is localized. The left-handed potential $V_L$ displays a shape of well, which tends to a double well as we modify the parameter $k_1$. Similar behavior also occurs for the scalar field profiles found in cubic and quartic gravity as seen in  Figs. (\ref{fig8})-(\ref{fig10}).

\begin{figure}[ht!]
\begin{center}
\begin{tabular}{ccc}
\includegraphics[scale=0.45]{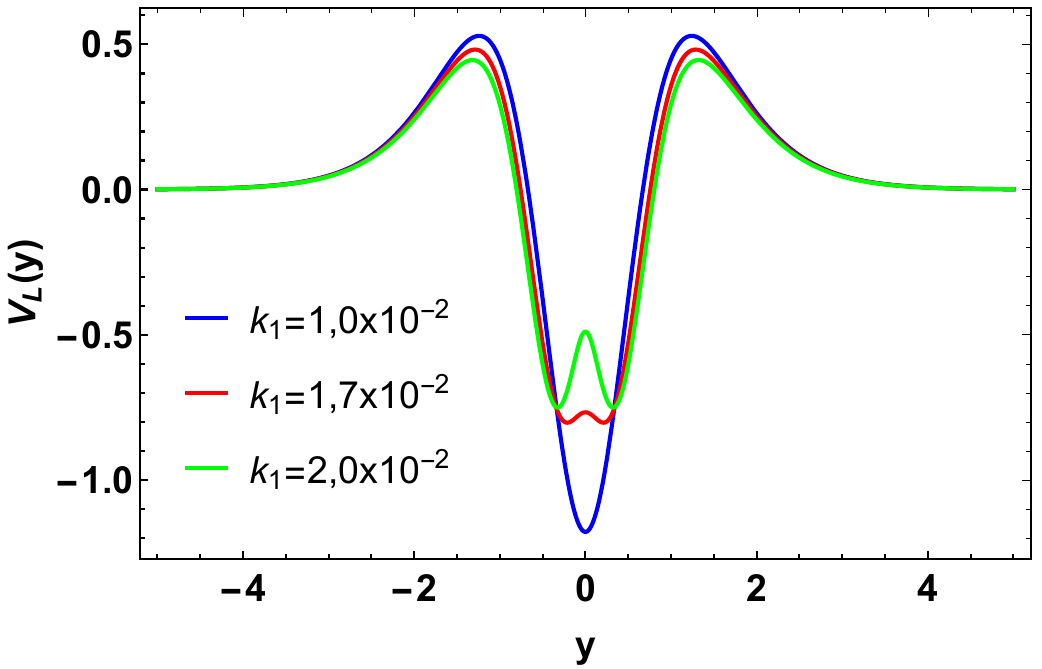} 
\includegraphics[scale=0.45]{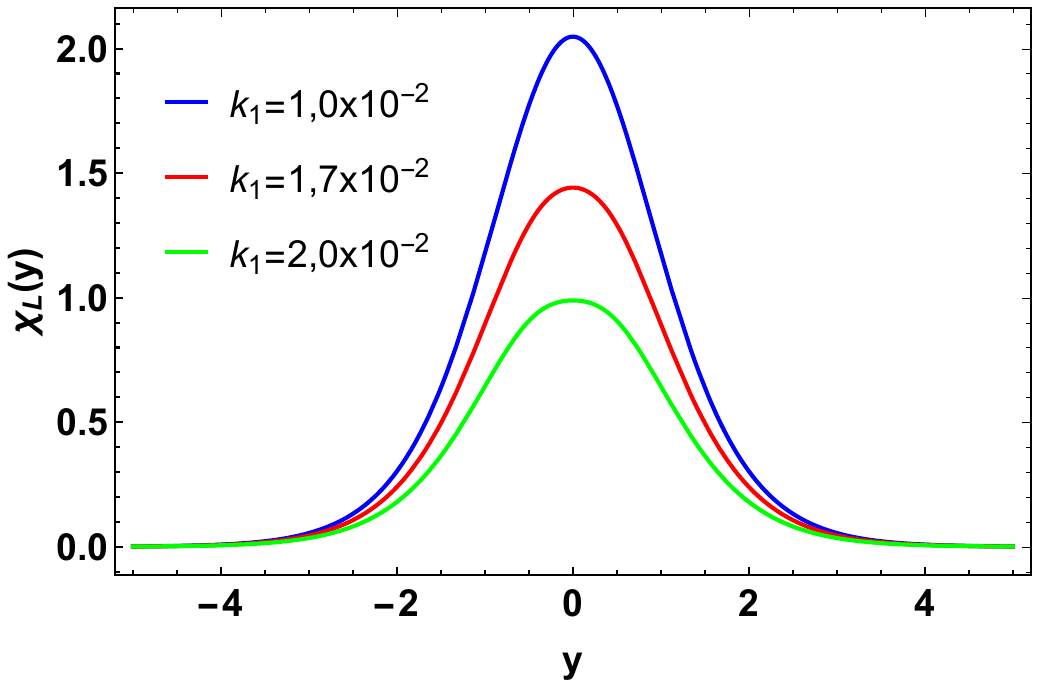}\\
(a)\hspace{6.7cm}(b)\\
\end{tabular}
\end{center}
\vspace{-0.5cm}
\caption{For the scalar field profile given by Eq. (\ref{sfeqq3}). (a) left-handed potential. (b) left-handed zero-mode. 
\label{fig11}}
\end{figure}
%%%%%%%%%%%%%%%%%%%%%%%%%%%%%%%%%%%%%%%%%%%%%%%%%%%%%%%%%%%%%%%%%%%%%%%%%%
\section{Rarita-Schwinger fermion localization}
\label{s5}

In this section, we investigate qualitatively the localization of spin $3/2$ massless fermions also known as gravitinos, which are described by a Rarita-Schwinger equation (for a review on gravitino, see e.g. \cite{VanNieuwenhuizen:1981ae}). Likewise spin $1/2$ fermion, the gravitino field requires a suitable coupling with the background scalar field \cite{Moreira:2023uys}. Then, we assume the 5D Rarita-Schwinger action \cite{Du:2015pjw,Zhou:2017xaq,Zhou:2023ejw}
\begin{align}\label{grav32}
    S_{\frac{3}{2}}=\int d^5x\sqrt{-g} (i\,\overline{\Psi}_M\Gamma^{[M} \Gamma^N \Gamma^{R]}\mathcal{D}_N \Psi_R-\lambda\phi\overline{\Psi}_M[\Gamma^{M},\Gamma^{N}]\Psi_N)
\end{align}
The covariant derivative now reads
\begin{align}\label{cds32}
\mathcal{D}_N \Psi_R=\partial_N \Psi_R+\Omega_N \Psi_R-\Gamma^{P}_{NR}\Psi_P    
\end{align}
Upon assuming the action (\ref{grav32}), one obtains the following equation of motion for gravitino
\begin{align}
i\Gamma^{[M} \Gamma^N \Gamma^{R]} \mathcal{D}_N \Psi_R-\lambda\phi[\Gamma^{M},\Gamma^{N}]\Psi_N=0    
\end{align}
Now with the metric (\ref{newmetric}) in hands, one has the nonvanishing components for the covariant derivative (\ref{cds32}) 
\begin{align}
&\mathcal{D}_\nu \Psi_\rho=\partial_\nu \Psi_\rho+\omega_\nu \Psi_\rho+\Dot{A}g_{\nu\rho}\Psi_4+\hf\Dot{A}\gamma_\nu\gamma_4 \Psi_\rho \\
&\mathcal{D}_4 \Psi_\nu=\partial_4 \Psi_\nu-\Dot{A}\Psi_\nu
\end{align}
It is convenient to choose the gauge condition $\Psi_4=0$. Thereby, we arrive at
\begin{align}\label{eqgravitino}
[i\gamma^{[\mu} \gamma^\nu \gamma^{\rho]} \partial_\nu + \gamma^\mu\gamma^4(\partial_z+\Dot{A})+\lambda\phi e^{A}[\gamma^\mu,\gamma^4]]\Psi_\rho(x,z) =0
\end{align}
Similarly to the above studied spin $1/2$ fermion case, we can write a KK chiral decomposition as follows
\begin{align}\label{kkgravitino}
\Psi_\mu (x,z)=\sum_n[\psi_{\mu L,n}(x)\alpha_{L,n}(z)+\psi_{\mu R,n}(x)\alpha_{R,n}(z)]e^{-A}   \end{align}
where the 4D section is constrained by the gauge condition $\partial^\mu\psi_\mu=\gamma^\mu\psi_\mu=0$. Besides, the 4D equations for gravitino look like
\begin{align}
i\gamma^{[\mu} \gamma^\nu \gamma^{\rho]} \partial_\mu\psi_{\nu L,n}=m[\gamma^\rho,\gamma^\alpha] \psi_{\alpha Rn} , \ i\gamma^{[\mu} \gamma^\nu \gamma^{\rho]} \partial_\mu\psi_{\nu R,n}=m[\gamma^\rho,\gamma^\alpha]\psi_{\alpha L,n},  \end{align}
and it obeys the following relation
\begin{align}
\gamma^4 \psi_{\mu L,n}=-\psi_{\mu L,n}    , \  \gamma^4 \psi_{\mu R,n}=\psi_{\mu R,n}.
\end{align}
Thus, by substituting the KK decomposition (\ref{kkgravitino}) into Eq.(\ref{eqgravitino}) and using the above properties for $\psi_{\mu L,n}$ and $\psi_{\mu R,n}$, we obtain the following set of coupled equations
\begin{align}
    &(\partial_z+\lambda\phi e^{A})\alpha_{R}=m\alpha_{L},\\
    &(\partial_z-\lambda\phi e^{A})\alpha_{L}=-m\alpha_{R},
\end{align}
which are rewritten as
\begin{align}
    &(-\partial_z^2+\mathcal{V}_L(z))\alpha_{L}=m^2\alpha_{L},\\
    &(-\partial_z^2+\mathcal{V}_R(z))\alpha_{R}=m^2\alpha_{R},
\end{align}
where we have defined the following effective potential for gravitino
\begin{align}
\mathcal{V}_L=(\lambda\phi e^{A})^2+\partial_z(\lambda\phi e^{A}),\\
\mathcal{V}_R=(\lambda\phi e^{A})^2-\partial_z(\lambda\phi e^{A}). 
\end{align}

At this point, it is worth observing the relation between the $\alpha$ and $\chi$, namely, $\alpha_R=e^{-A}\chi_L$. Such a change promotes a small increase in the amplitude of the zero-mode for the spin 3/2. Therefore, the behavior of the zero-mode for the gravitino is similar to those already demonstrated for the Dirac fermions. Unlike to spin 1/2 fermions, only the right-handed zero-mode for gravitino is localized on the brane as addressed previously \cite{Du:2015pjw, Zhou:2017xaq, Zhou:2023ejw}. On the other hand, as we can see the effective potentials are the same for both fields with opposing chiralities. In this sense, it would be repetitive to graphically represent such solutions.

%%%%%%%%%%%%%%%%%%%%%%%%%%%%%%%%%%%%%%%%%%%%%%%%%%%%%%%%%%%%%%%%%%%%%%%%%%%
\section{Final remarks}
\label{s6}

In this paper, we studied a five-dimensional braneworld sourced by a single scalar field in the $f(R,Q,P)$ modified gravity. We have considered a specific form for warp factor to obtain a complete description of braneworld. We obtained the matter field with deformed profile by modifying the parameters that control the particular gravity model considered in this work. To avoid any instability problems we have treated them in the context of effective field theory.

As concrete models of $f(R,Q,P)$ gravity, one chooses proper functions corresponding to quadratic, cubic, and quartic gravities. For each one of them, we have made a suitable choice of the parameters so that we can generate deformed kink-like structures. For instance, in the case of quadratic gravity, a two-kink solution was obtained, leading to the phenomena of brane splitting. Such phenomena can be seen as a phase transition that the matter field suffers as it nears the brane core. Additionally, in the case of cubic and quartic gravities, it was possible to observe the emergence of 3-kink solutions, suggesting that higher-order invariants provoke a richer internal structure. Moreover, upon examining the energy density, one noted that internal structure appears right at the brane core (the center of the energy density) in the case of quadratic and quartic gravities. On the other hand, the cubic gravity leads to a slight modification in the lateral region of the energy density as we can observe in Fig.(\ref{fig3})(d).

Another issue we investigated was the localization of fermionic fields (Dirac fermions and gravitinos) using Yukawa-like coupling with the scalar field.  In particular, we have shown that only the left-handed zero-mode for spin 1/2 fermions is localized on the brane as it occurs in the context of thick branes in GR. On the other hand, for the case of spin 3/2 fermions, only the right-handed zero-mode is localized on the brane. The localization of fermions is pivotal to validate the physical consistency of our model.

We could extend this work with further papers by investigating the stability of brane under small tensor perturbations. In this sense, it is possible to study how the graviton massive modes would  modify the Newton gravity law. Moreover, the localization of abelian gauge fields could be analyzed. Another issue we could investigate is the phase transitions of field configuration utilizing the configurational entropy framework. We plan to address these issues in forthcoming papers.

\section*{Acknowledgments}
\hspace{0.5cm} The authors thank the Funda\c{c}\~{a}o Cearense de Apoio ao Desenvolvimento Cient\'{i}fico e Tecnol\'{o}gico (FUNCAP), the Coordena\c{c}\~{a}o de Aperfei\c{c}oamento de Pessoal de N\'{i}vel Superior (CAPES), Paraiba State Research Foundation (FAPESQ-PB) and the Conselho Nacional de Desenvolvimento Cient\'{i}fico e Tecnol\'{o}gico (CNPq).  Fernando M. Belchior thanks the Departmento de F\'isica da Universidade Federal da Para\'iba - UFPB for the kind hospitality and has been partially supported by CNPq grant No. 161092/2021-7. Roberto V. Maluf thanks the CNPq for grant no. 200879/2022-7. Albert Yu. Petrov
 thanks the Brazilian agency FAPESQ-PB (process No. 150891/2023-7) and CNPq (grant No. 303777/2023-0).  Paulo J. Porf\'irio thanks the Brazilian agency FAPESQ-PB (process No. 150891/2023-7) and CNPq (grant No. 307628/2022-1).

\end{document}